\documentclass[journal, a4paper]{IEEEtran}

\usepackage{amsmath,amssymb,amsfonts,amsthm,mathtools}

\usepackage[utf8]{inputenc}
\usepackage[T1]{fontenc}
\usepackage{hyperref}
\usepackage{url}
\usepackage{booktabs}
\usepackage{nicefrac}
\usepackage{microtype}
\usepackage{xcolor}
\usepackage{graphicx}
\usepackage{subfigure}
\usepackage{multirow}
\usepackage{cite}
\usepackage{bm}
\usepackage{cases}
\usepackage{float}
\usepackage{stfloats}
\usepackage{algorithm}
\usepackage{algorithmic}

\newtheorem{theorem}{{Theorem}}

\newtheorem{remark}{{Remark}}

\begin{document}
\title{From Interference to Diversity: Mode Coloring and\\Two-Dimensional Modal-Time Rake Combining for Multipath OAM Communications}


\author{%
\IEEEauthorblockN{Runyu Lyu, \textit{Member, IEEE}}
\thanks{Runyu Lyu is with the School of Electronics and Information Engineering, Nanjing University of Information Science and Technology, Nanjing, 210044, China (e-mail: runyulyu@nuist.edu.cn).}
}

\IEEEtitleabstractindextext{%
\begin{abstract}
Orbital angular momentum (OAM) beams provide a spatial-multiplexing dimension for wireless communications, but multipath scattering broadens their mode spectrum, so that energy leaks across modes as inter-mode interference (IMI), which conventional OAM demultiplexing discards. This paper treats the leaked modal energy not as interference but as a structured, harvestable resource: it attains up to $LK$-fold asymptotic diversity under independent-branch fading and yields harvested energy under the rank-limited per-tap fading of the operating channel. Each active OAM mode is colored with a distinct time-domain orthogonal spreading code, and a two-dimensional (2D) receiver grid spanning the OAM modes and the resolvable multipath delays jointly combines the despread outputs. The design builds on the Bessel order-difference coupling law of tilted OAM beams, summarized by a Gaussian mode-diffusion model with a mode-coherence metric that quantifies the surviving orthogonality, and on an effective diversity metric that separates the combining gain available at finite signal-to-noise ratio (SNR) from the asymptotic diversity order. At 28~GHz, with a four-mode eight-element uniform circular array (UCA) and four resolvable multipath taps, the proposed 2D modal-time Rake lowers the bit-error rate (BER) by one to three orders of magnitude and saves roughly 10--15~dB of required SNR over mode-wise detection and mode-domain zero-forcing (ZF) and minimum mean-square error (MMSE) detection, and it also outperforms an uncolored modal-time MMSE equalizer; the advantage survives the spectral-efficiency cost of spreading as well as channel-estimation and timing imperfections.
\end{abstract}

}

\maketitle
\IEEEdisplaynontitleabstractindextext
\IEEEpeerreviewmaketitle

\section{Introduction}
\IEEEPARstart{W}{ith} conventional spectral resources saturating, new degrees of freedom of electromagnetic waves are a central route to higher spectral efficiency. Orbital angular momentum (OAM) is one of them: an OAM wave has a helical phase front $\exp(j\ell\varphi)$, where the topological charge $\ell$ is an integer, and waves with different charges are mutually orthogonal \cite{allen1992orbital, thide2007utilization}. OAM thus offers a mode-division-multiplexing (MDM) dimension that is independent of frequency, time, and polarization \cite{cheng2019orbital, chen2020multimode, willner2015optical}. Since the first radio-domain proposals \cite{thide2007utilization, mohammadi2010orbital} and the multi-gigabit-per-second millimeter-wave experiments \cite{tamburini2012encoding, yan2014high}, OAM has been studied intensively for 6G and beyond.

At radio frequencies, the standard OAM transceiver employs a uniform circular array (UCA): an $N$-element UCA fed with the progressive azimuthal phase $\exp(j\ell\varphi_n)$ radiates mode $\ell$, and an $N$-point discrete Fourier transform (DFT) over the received element signals recovers the modal content \cite{chen2020multimode, mohammadi2010orbital, edfors2011unexploited}. Under ideal line-of-sight (LoS) conditions the modal channels are orthogonal and the DFT receiver is optimal.

In practice, the received field superposes a LoS component with many reflected and scattered components that carry different delays, attenuations, and angles of arrival. Two effects then degrade the modal orthogonality. First, scattering broadens the spiral spectrum: the azimuthal structure of each OAM beam is perturbed, so energy of mode $\ell$ leaks into neighboring modes $\ell'$; this inter-mode interference (IMI) has been measured in millimeter-wave indoor links \cite{yan2015experimental, yan2016multipath} and in 24--30~GHz campaigns \cite{liao2022indoor, wang2023propagation, hu2025fr3}, and channel models and terahertz inter-satellite studies confirm that it is a first-order impairment of OAM-MDM \cite{chen2020multipathbook, lyu2025gbsm, gao2025intersatellite}. Second, the delay spread splits the signal into $K$ resolvable taps, so a single-shot DFT demultiplexing also mixes modal and temporal information.

A conventional DFT demultiplexer treats IMI as additive distortion: it projects the field onto each mode and discards everything outside it. Its bit-error rate (BER) is then dominated by an interference floor that does not vanish with transmit power, and the scattered energy is lost.

Multipath is not the only threat: in free space the OAM beam also suffers an intrinsic hollow divergence that widens with $|\ell|$ \cite{zheng2015twisted}, which can be alleviated at the transmitter by beam shaping, constant-beam-radius arrays, and fractal launchers \cite{wang2024constantradius, cheng2024fractal}. Those transmit-side techniques and the receiver-side scheme proposed here are complementary: beam shaping keeps the wavefront intact, whereas residual-orthogonality extraction harvests whatever orthogonality survives the channel \cite{lyu2026eit}.

Prior work on robust OAM reception falls into two groups. Antenna-domain mitigation, from Cassegrain-aperture backhaul links \cite{wu2025cassegrain} to constant-beam-radius and fractal launchers \cite{wang2024constantradius, cheng2024fractal}, improves the spatial confinement and the received power of the beam, rather than explicitly exploiting multipath-induced modal leakage. Signal-processing schemes combine OAM with orthogonal frequency-division multiplexing (OFDM) and phase compensation for sparse multipath \cite{liang2020joint} and with multiple-input multiple-output (MIMO) equalization of misalignment-induced crosstalk \cite{icoch2016mimo}; they combine modal branches without a delay dimension, and none uses per-mode spreading signatures, whereas fiber OAM-MDM applies $L\times L$ MIMO digital signal processing to deterministic, delay-free mode coupling \cite{bozinovic2013terabit}. Code-domain schemes use OAM with optical code-division multiple access (CDMA) in free-space optics \cite{singh2022oamocdma} and with spread-spectrum signaling for security \cite{zhang2020spread}, but for multiple access or confidentiality rather than to restore multipath-destroyed orthogonality. OAM systems assisted by reconfigurable intelligent surfaces (RISs) have also been studied \cite{lyu2022risswipt}. On the theoretical side, the spiral-spectrum description \cite{torner2005spiral, molina2002management}, the Gaussian broadening of turbulence-disturbed OAM spectra \cite{paterson2005turbulence}, and the misaligned-beam analyses of \cite{vasnetsov2005misaligned, nieminen2025misaligned} established that the coupling of a tilted OAM beam depends on the mode-order \emph{difference} $\ell'-\ell$, the backbone of the channel model of Section~II.

In ideal LoS, UCA-based OAM-MDM is equivalent to conventional MIMO spatial multiplexing, with the DFT acting as a spatial pre/post-coder \cite{edfors2011unexploited, opare2015mode}. The distinction appears in multipath: generic MIMO processing learns an unstructured channel matrix, which zero-forcing (ZF) or minimum mean-square error (MMSE) array-domain detectors must invert, whereas the mode-domain view exposes a \emph{structured} coupling, a mode-order-difference kernel that produces IMI convolved with a delay spread that produces inter-symbol interference (ISI), which the mode coloring and the $L\times K$ finger grid exploit by direct, non-iterative linear processing.

We draw a signal-processing \emph{correspondence}, not an isomorphism, between the OAM multipath problem and the classical CDMA multipath problem (Table~\ref{tab:iso}), because the two stochastic channel models are not mathematically isomorphic. In CDMA each user carries a unique spreading code, and the two-dimensional (2D) Rake receiver \cite{zoltowski2000reduced, ramos1997blind} collects the energy dispersed over the antenna array and the delay taps by maximum-ratio combining (MRC) or by MMSE combining. Mapping OAM modes to CDMA users and IMI to multiple-access interference (MAI), the proposed receiver borrows the 2D-Rake principle and adapts it to the mode-delay structure of OAM multipath channels.

\begin{table}[t]
\centering
\caption{The CDMA-to-OAM signal-processing correspondence exploited in this paper.}
\label{tab:iso}
\begin{tabular}{@{}c c c@{}}
\toprule
& CDMA 2D-Rake & OAM modal-time 2D-Rake (proposed)\\
\midrule
Dimension 1 & antenna element & OAM mode $\ell$\\
Dimension 2 & multipath delay & multipath delay\\
Spreading & user code & per-mode orthogonal code\\
Finger & (element, delay) & (mode, delay)\\
Interference & MAI & IMI\\
Combining & MRC/MMSE & MRC/MMSE\\
\bottomrule
\end{tabular}
\end{table}

Based on this correspondence, the contributions are as follows.
\begin{enumerate}
\item \textbf{Mode coloring.} Each active OAM mode carries a distinct time-domain orthogonal spreading code, a scheme called \emph{time-domain orthogonal code mode coloring} (TDOCC), which gives the modes distinguishable temporal signatures and lets the receiver recover energy that would otherwise be treated as interference.
\item \textbf{Validated channel model.} The per-ray coupling of a tilted OAM beam is the Bessel function $J_{\ell'-\ell}(kR\sin\beta)$ of the mode-order \emph{difference} and follows from the Jacobi--Anger expansion; its angular-spread average motivates a tractable Gaussian mode-diffusion model of the IMI. Both are validated in terms of the normalized mean-square error (NMSE) against free-space UCA field simulation.
\item \textbf{Mode coherence.} The metric $\rho$, the ratio of diagonal to off-diagonal coupling power, quantifies the residual orthogonality; it is an interference indicator, not a diversity sufficient statistic.
\item \textbf{Two-dimensional modal-time Rake receiver.} An $L\times K$ finger grid with MRC and joint MMSE combining suppresses IMI and ISI; the diversity order is $LK$ under independent-branch conditions and is bounded by the modal covariance rank otherwise (Section~IV).
\item \textbf{Diversity and spectral-efficiency analysis.} The asymptotic diversity order is separated from the effective metric that governs finite signal-to-noise ratio (SNR) behavior, the condition under which IMI becomes diversity is identified, and the spectral-efficiency cost of coloring is quantified under an explicit normalization.
\end{enumerate}

To the best of our knowledge, we are unaware of prior wireless OAM work that jointly exploits per-mode temporal orthogonal signatures and modal-delay combining to recover multipath-induced modal leakage as a diversity resource: code-division OAM exists for optical access \cite{singh2022oamocdma} and for security \cite{zhang2020spread}, and modal diversity has been explored in optical OAM links under turbulence \cite{willner2015optical}.

Section~II presents the system model and the transceiver architecture. Section~III develops the mode-time channel model and the proposed 2D-Rake receiver. Section~IV analyzes the diversity order and the spectral-efficiency trade-off. Section~V reports the numerical results. Section~VI concludes the paper.

\section{System Model}
\label{sec:system}

\begin{figure}[b]
\centering
\subfigure[System model.]{\includegraphics[width=0.88\columnwidth]{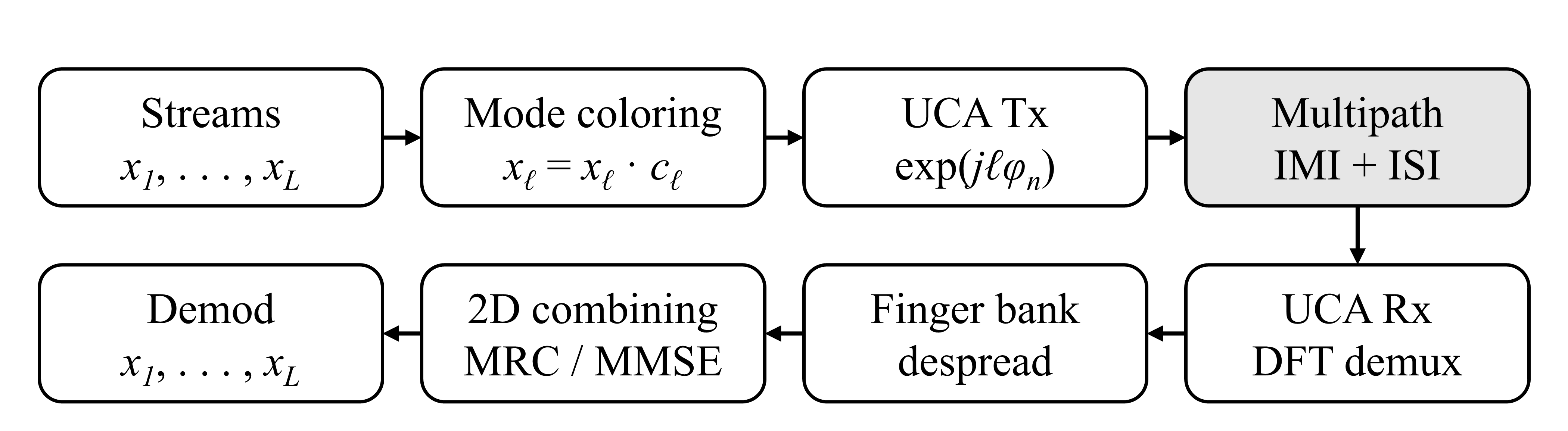}}
\subfigure[Transceiver architecture.]{\includegraphics[width=0.88\columnwidth]{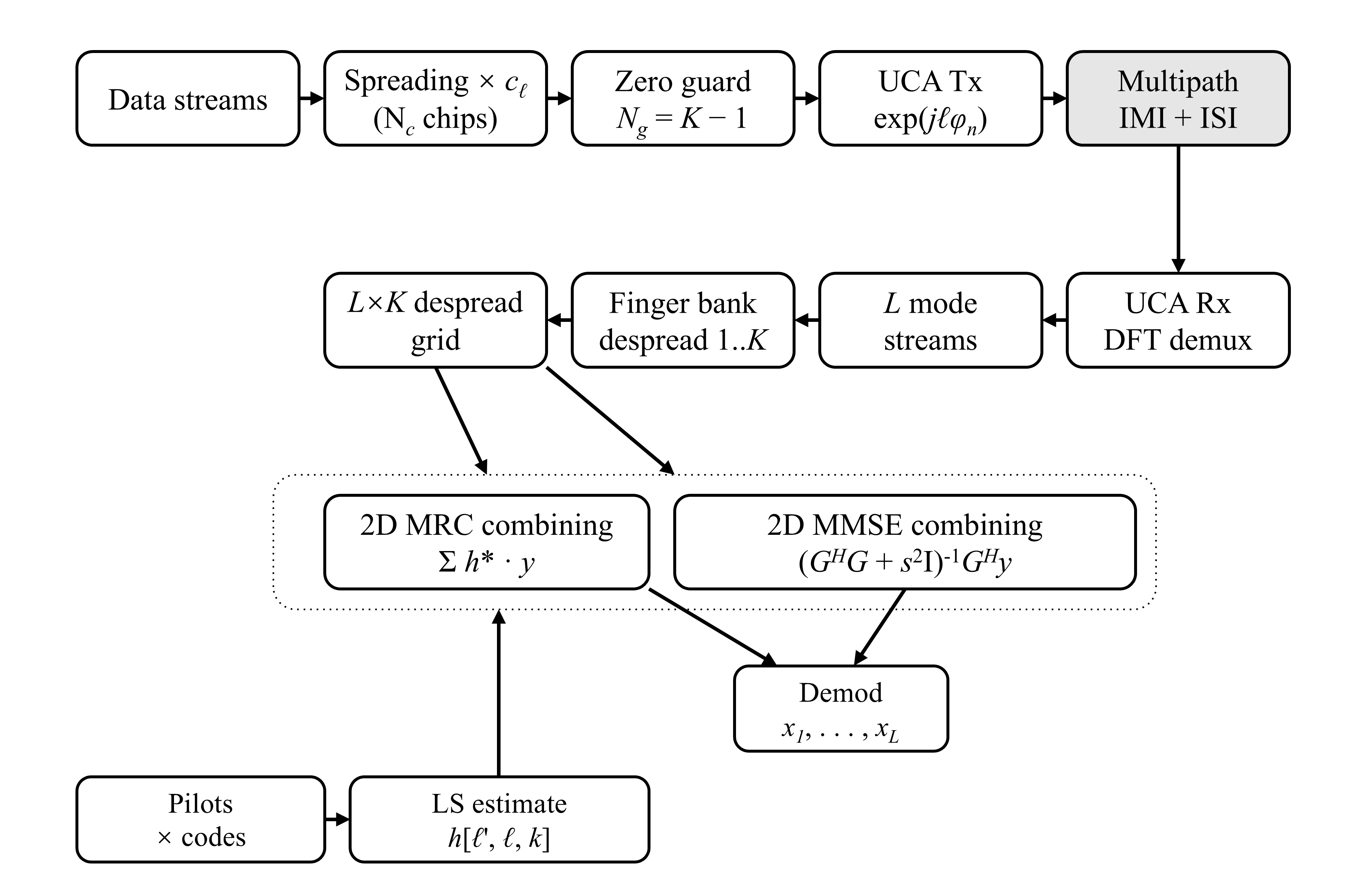}}
\caption{Proposed TDOCC + 2D-Rake transceiver.}
\label{fig:system}
\end{figure}

We consider a point-to-point OAM link in which the transmitter and the receiver each employ an $N$-element UCA of radius $R$, and $L$ OAM modes are active. In a multipath channel the received field contains a LoS ray and $K$ resolvable scattered rays, so two impairments appear at once: scattering leaks the energy of each mode into neighboring modes, the IMI that degrades modal orthogonality, and the delay spread distributes each symbol over the $K$ rays, the ISI that mixes consecutive symbols.

Fig.~\ref{fig:system}(a) shows the resulting transceiver. At the transmitter, the $L$ mode streams undergo TDOCC: mode $\ell$ is multiplied by its own orthogonal code of length $N_c$ chips, and a zero guard interval of $N_g$ chips follows each symbol so that its multipath spread does not overlap the next symbol. The colored streams are fed to the transmit UCA with the progressive phase of mode $\ell$ and radiated as a superposition. At the receiver, the array outputs are DFT-demultiplexed into $L$ modal streams; each stream is despread with each mode's code at each of the $K$ delay taps, which forms an $L\times K$ finger bank; and the finger outputs are combined either by MRC or by a joint modal-time MMSE equalizer that also orthogonalizes the residual IMI and ISI. Fig.~\ref{fig:system}(b) details this flow, including the pilot-based channel estimation.

Throughout, ``modal-time 2D-Rake architecture'' denotes the receiver: the \emph{2D modal-time MRC Rake} combines the fingers conjugate-wise, whereas the \emph{joint modal-time MMSE equalizer} combines them by a matrix operation and is not a per-finger Rake in the strict sense. Unless stated otherwise the receiver has perfect channel state information (CSI); imperfect CSI is quantified in Section~\ref{sec:csi}. Section~III develops the mode-time channel model and then the two combiners.

\section{Mode-Time Channel Model and Proposed 2D-Rake Receiver}
\label{sec:proposed}

\subsection{UCA transceiver and per-ray mode-coupling law}
We consider a point-to-point link whose transmitter and receiver are $N$-element UCAs of radius $R$ on the common $z$-axis, separated by a distance $D\gg R$; element $n$ sits at azimuth $\varphi_n=2\pi n/N$. To generate the OAM mode $\ell$, the transmitter feeds element $n$ with the progressive phase
\begin{equation}
s_n^{(\ell)} = \exp\!\big(j\,\ell\,\varphi_n\big).
\label{eq:txphase}
\end{equation}
With an isotropic radiating element, the far-field of the array is
\begin{equation}
E(r,\theta,\varphi) \;\propto\; \frac{e^{-jkr}}{r}\sum_{n=0}^{N-1} s_n^{(\ell)}\, e^{\,j k R \sin\theta \cos(\varphi-\varphi_n)},
\label{eq:arrayfactor}
\end{equation}
where $(r,\theta,\varphi)$ are the spherical coordinates of the observation point, $k=2\pi/\lambda$ is the wavenumber, and $\lambda$ is the wavelength. The Jacobi--Anger expansion together with the azimuthal orthogonality of the array \cite{thide2007utilization} reduces \eqref{eq:arrayfactor}, for $|\ell|<N/2$ and moderate $kR\sin\theta$, to a single Bessel term,
\begin{equation}
E(r,\theta,\varphi) \;\approx\; A\,j^{\ell}\,J_{\ell}(kR\sin\theta)\,e^{j\ell\varphi}\,\frac{e^{-jkr}}{r},
\label{eq:besselbeam}
\end{equation}
where $A$ is a constant amplitude; the resulting field is the familiar Bessel-form OAM beam, whose azimuthal index is the OAM charge and whose radial profile is $J_\ell$. With the $\lambda/2$-spacing convention $R=N\lambda/(4\pi)$, modes $|\ell|<N/2$ are generated unambiguously. At the receiver the modal content is recovered by an $N$-point DFT,
\begin{equation}
r_{\ell'} = \frac{1}{N}\sum_{n=0}^{N-1} r_n\, e^{-j\ell'\varphi_n},
\label{eq:dft}
\end{equation}
which under ideal LoS projects mode $\ell$ exactly onto $\ell'=\ell$.

Consider a ray of mode $\ell$ arriving with polar angle $\beta$ relative to the array normal, a reflection-induced tilt. Across the small receive aperture the tilted OAM beam is approximated by
\begin{equation}
E(\psi) \;=\; A\,e^{j\ell \psi}\, e^{-j k R \sin\beta \cos(\psi-\phi_s)},
\label{eq:tilted}
\end{equation}
where $\psi$ is the azimuth on the receive array and $\phi_s$ the ray azimuth; the first factor is the helical phase, the second the plane-wave tilt factor. Projecting \eqref{eq:tilted} onto the receive OAM basis with \eqref{eq:dft} and expanding the tilt factor with the Jacobi--Anger expansion, only the term with $m=\ell'-\ell$ survives the azimuthal sum, giving
\begin{equation}
r_{\ell'} \;=\; A\,(-j)^{\ell'-\ell} J_{\ell'-\ell}\big(kR\sin\beta\big)\, e^{-j(\ell'-\ell)\phi_s},
\label{eq:couplinglaw}
\end{equation}
so that the normalized magnitude of the coupling coefficient $c_{\ell'}$ to output mode $\ell'$ follows the \emph{Bessel order-difference law}
\begin{equation}
\frac{|c_{\ell'}|}{|c_{\ell}|} = \frac{\big|J_{\ell'-\ell}(kR\sin\beta)\big|}{|J_0(kR\sin\beta)|}, \qquad \lim_{\beta\to0}|c_{\ell'}| = \delta_{\ell',\ell}\, |c_\ell| .
\label{eq:bessel_law}
\end{equation}
Thus (i) a perfectly aligned ray ($\beta=0$) is diagonal, (ii) the coupling decays with the mode-order difference $d=\ell'-\ell$, and (iii) increasing tilt broadens the modal coupling distribution; Section~V-B validates the law against full free-space UCA field simulation. The order-difference structure is exactly that established for tilted/displaced OAM beams in optics \cite{vasnetsov2005misaligned, nieminen2025misaligned}; the specific argument $kR\sin\beta$ follows from the Jacobi--Anger derivation above.

\subsection{Angular-spread average: Gaussian mode diffusion}
A multipath cluster is a family of rays whose tilt angles $\beta$ have root-mean-square (RMS) width $\Delta\theta$. Averaging the squared coupling law \eqref{eq:bessel_law} over this distribution concentrates the coupling around $d=0$, with a width set by the angular spread. Disturbed OAM beams are commonly modeled as acquiring an approximately Gaussian broadening of their mode spectrum, as in the turbulence analyses of \cite{paterson2005turbulence, rodenburg2012turbulence}, which also note that the exact mode-dependent distribution is not strictly Gaussian. This observation \emph{motivates} the tractable Gaussian mode-diffusion \emph{amplitude} kernel
\begin{equation}
g_{d} \;=\; \exp\!\Big(-\frac{d^{2}}{2\sigma_\ell^{2}}\Big),
\qquad \sigma_\ell \;=\; \frac{kR\sin(\Delta\theta)}{\sqrt{2}},
\label{eq:diffusion}
\end{equation}
where $\sigma_\ell$ is the mode-domain diffusion bandwidth. Equation~\eqref{eq:diffusion} is a modeling assumption, not an exact law: Section~V-B verifies that it reproduces the per-ray Bessel law and its angular average in the operating regime. Note that $\sigma_\ell\propto kR$: a larger array resolves more modes but is proportionally more sensitive to IMI.

\subsection{Mode-time channel and mode coherence}
With $K$ resolvable delay taps, the complete channel is described by the per-tap coupling kernels $h[\ell',\ell,k]$, the coupling from input mode $\ell$ at delay $k$ to output mode $\ell'$. We collect the kernels in a block channel matrix $\mathbf{H}\in\mathbb{C}^{LK\times LK}$ whose $(k',k)$-th $L\times L$ block is
\begin{equation}
[\mathbf{H}]_{k',k} = \begin{cases}
\mathbf{H}_k, & k'=k,\\
\mathbf{0}, & k'\neq k,
\end{cases}
\label{eq:Hblock}
\end{equation}
where $\mathbf{H}_k$ is the tap-$k$ mode-coupling matrix, whose diagonal holds the self-mode gains and whose off-diagonal entries follow \eqref{eq:diffusion} and carry the azimuth phase of \eqref{eq:couplinglaw}. The LoS tap ($k=1$) is exactly diagonal, while the scattered taps $k=2,\dots,K$ follow the exponential power-delay profile (PDP)
\begin{equation}
P_k = \frac{e^{-(k-1) T_c/\tau_{d}}}{\sum_{m=0}^{K-1} e^{-m T_c/\tau_{d}}},\qquad k=1,\dots,K,
\label{eq:pdp}
\end{equation}
where $P_k$ is the mean power of tap $k$ and $\tau_{d}$ is the exponential decay constant of the profile, not the RMS delay spread: the simulation value $\tau_d=2T_c$ corresponds to an actual RMS delay spread of $\tau_{\mathrm{rms}}\approx1.0\,T_c$, computed from the second central moment of the discrete profile.

To quantify the surviving orthogonality, we define the \emph{mode coherence}
\begin{equation}
\rho \;=\; \frac{\sum_{\ell} \big|[\mathbf{C}]_{\ell,\ell}\big|^{2}}
{\sum_{\ell\neq\ell'} \big|[\mathbf{C}]_{\ell,\ell'}\big|^{2}},
\qquad \mathbf{C}=\sum_{k}\mathbf{H}_{k}\,
\label{eq:coherence}
\end{equation}
the ratio of diagonal to off-diagonal coupling power of the total mode-coupling matrix $\mathbf{C}$. Perfect LoS gives $\rho\to\infty$, whereas $\rho\approx1$ means that the mode domain is essentially mixed. The denominator is the \emph{total} (not per-pair-averaged) cross-mode power, so $\rho$ depends on the active mode count $L$; moreover, $\rho$ is an interference indicator, not a sufficient statistic for diversity, since two channels with the same $\rho$ can have very different modal covariance ranks (Section~\ref{sec:divC}). Section~V-C shows that $\rho$ decreases monotonically with $\Delta\theta$ yet stays above unity over a broad range, which defines the regime where residual-orthogonality recovery pays off.

\subsection{Mode-colored transmission and received signal}
The transmission format follows the TDOCC principle: each of the $L$ active modes is assigned an orthonormal spreading code $\mathbf{c}_{\ell}\in\mathbb{C}^{N_c}$, $\ell=1,\dots,L$, drawn from a Walsh--Hadamard matrix so that
\begin{equation}
\mathbf{c}_{\ell}^{H}\mathbf{c}_{\ell'} = \delta_{\ell,\ell'}.
\label{eq:orthcode}
\end{equation}
which requires $N_c\ge L$; for the Walsh--Hadamard family $N_c$ is restricted to powers of two. A quadrature phase-shift keying (QPSK) symbol $x_\ell[s]$ on mode $\ell$ is spread over $N_c$ chips \emph{within} the symbol period,
\begin{equation}
\bar{x}_\ell[sN_c+n] = x_\ell[s]\,c_\ell[n], \qquad n=0,\dots,N_c-1,
\label{eq:spread}
\end{equation}
and a \emph{zero-guard interval} of $N_g=K-1$ chips is inserted between consecutive symbols so that the multipath spread of one symbol does not overlap its successor. The guard length satisfies
\begin{equation}
N_g \;\ge\; \Big\lceil \frac{\tau_{\max}}{T_c} \Big\rceil
\label{eq:guard}
\end{equation}
under the integer-delay (chip-resolution) channel model, where $\tau_{\max}$ is the maximum excess delay and $T_c$ the chip duration. Fractional delays violate exact chip-level isolation; their effect is quantified in Section~\ref{sec:timing}. The colored signal of each mode is then fed to the UCA with the phase profile \eqref{eq:txphase}; the $n$-th element transmits the superposition of all colored modes. In the resulting frame, the $L$ codes are the columns of the Walsh--Hadamard matrix, $x_\ell[s]\mathbf{c}_\ell$ occupies $N_c$ chips, and the following $N_g$ zero chips are long enough that the $K$ multipath replicas of a symbol all arrive inside the guard and leave symbol $s+1$ undisturbed.

After propagation through the mode-time channel and DFT demultiplexing \eqref{eq:dft}, the chip-domain signal at output mode $\ell'$ is
\begin{equation}
r_{\ell'}[m] = \sum_{\ell=1}^{L}\sum_{k=1}^{K}
h[\ell',\ell,k]\;\bar{x}_{\ell}[m-k+1]
+ n_{\ell'}[m],
\label{eq:rx}
\end{equation}
where $n_{\ell'}[m]\sim\mathcal{CN}(0,\sigma_n^2)$ is additive white Gaussian noise. Equation~\eqref{eq:rx} is the starting point of the receiver design.

\subsection{Channel estimation and pilot structure}
\label{sec:ce}
The MMSE combiner requires the coupling tensor $\{h[\ell',\ell,k]\}$. Because the colored modes already carry orthogonal signatures, the pilot design is natural: in a block-pilot phase, each mode transmits a known pilot symbol spread with its own code, and the receiver despreads the $L$ streams at each of the $K$ delays with the code of the piloting mode. The resulting least-squares (LS) estimate of tap $(\ell',\ell,k)$ is
\begin{equation}
\hat{h}[\ell',\ell,k] = h[\ell',\ell,k] + e[\ell',\ell,k],
\qquad e \sim \mathcal{CN}\!\Big(0, \frac{\sigma_n^{2}}{E_p}\Big),
\label{eq:ls}
\end{equation}
where, with the normalized codes of \eqref{eq:orthcode}, $E_p$ is the energy of the pilot \emph{symbol}, because the zero-lag orthogonality of the codes separates the modes and the guard separates the taps. The relative estimation error $\epsilon=\sigma_n/\sqrt{E_p}/\|h\|$ is the quantity swept in Section~\ref{sec:csi}. With one pilot symbol per mode per coherence block the pilot overhead is $L/N_{\mathrm{sym,coh}}$, where $N_{\mathrm{sym,coh}}$ is the number of information symbols per coherence block, i.e., $4\%$ for a hundred-symbol block. Coloring also makes the mode-time channel estimable with a single-shot correlator, whereas an uncolored system must solve a joint estimation and detection problem.

\subsection{Two-dimensional finger bank}
For the symbol occupying chips $[sN_c, sN_c+N_c-1]$, the receiver correlates every output-mode stream with every mode's code at every delay tap. Finger $(\ell_{\mathrm{ref}},\ell',k)$ despreads the stream of output mode $\ell'$ with the code of the reference (desired) mode $\ell_{\mathrm{ref}}$ at delay $k$:
\begin{equation}
y[\ell_{\mathrm{ref}};\ell',k] = \sum_{n=0}^{N_c-1}
r_{\ell'}\!\big[s N_c + (k-1) + n\big]\;c_{\ell_{\mathrm{ref}}}^{*}[n].
\label{eq:finger}
\end{equation}
Two properties make the finger a modal Rake. First, because the codes are orthogonal at zero lag \eqref{eq:orthcode}, the finger with $\ell_{\mathrm{ref}}=\ell$ isolates the contribution of mode $\ell$ regardless of the IMI, since the same-delay cross terms $\propto\mathbf{c}_{\ell}^{H}\mathbf{c}_{\ell'}$, $\ell'\neq\ell$, vanish in \eqref{eq:rx}. Second, the delay offset $k$ aligns the finger with the $k$-th resolvable tap, so it also collects temporal diversity. The complete finger output for mode $\ell$ is the $L\times K$ matrix
\begin{equation}
\mathbf{Y}_\ell[s] = \big[y[\ell;\ell',k]\big]_{\ell'=1,\dots,L;\, k=1,\dots,K},
\label{eq:fingermatrix}
\end{equation}
whose entries are the estimates of $x_\ell[s]$ observed on each output mode and each delay.

\subsection{MRC combining}
The direct modal-domain adaptation of the CDMA 2D-Rake is MRC: the despread values are weighted by the conjugates of the corresponding channel couplings and summed,
\begin{equation}
\hat{x}_{\ell}[s] =
\frac{\sum_{\ell'=1}^{L}\sum_{k=1}^{K} h^{*}[\ell',\ell,k]\; y[\ell;\ell',k]}
{\sqrt{\sum_{\ell',k}\big|h[\ell',\ell,k]\big|^{2}}}.
\label{eq:mrc}
\end{equation}
Crucially, the summation runs over \emph{all} output modes $\ell'$, not only $\ell'=\ell$: energy of mode $\ell$ that leaked into other modes, which a conventional DFT demultiplexer discards as IMI, is harvested coherently. In the idealized case of $LK$ independent Rayleigh-faded fingers with equal power, \eqref{eq:mrc} achieves the diversity order $LK$ (Section~\ref{sec:divA}); in interference-limited regimes, the residual IMI/ISI at non-zero code lags saturates MRC, motivating the MMSE combiner below.

\subsection{MMSE combining}
To suppress the residual IMI and ISI, we jointly equalize the mode-time grid. Concatenating the $L$ output-mode streams of one symbol into a single observation over the window of $W=N_c+N_g$ chips,
\begin{equation}
\mathbf{y}[s] = \mathrm{vec}\big(r_{\ell'}[sN_c+1:sN_c+W]\big) \in \mathbb{C}^{LW},
\label{eq:obs}
\end{equation}
the symbol-level model reads
\begin{equation}
\mathbf{y}[s] = \mathbf{G}\,\mathbf{x}[s] + \mathbf{n},
\label{eq:mmse_model}
\end{equation}
where $\mathbf{x}[s]=[x_1,\dots,x_L]^{T}$ and $\mathbf{G}\in\mathbb{C}^{LW\times L}$ is the effective symbol-level channel matrix: its $m$-th column is the response of a unit symbol on mode $m$ to the observation \eqref{eq:obs}, i.e., the convolution of the spreading code $\mathbf{c}_m$ with the mode-time channel (obtained in closed form, or, as in our simulations, by passing unit symbols through the channel). With the Gaussian observation model, the linear minimum-mean-square-error estimate of the symbol vector is
\begin{equation}
\hat{\mathbf{x}}[s] = \big(\mathbf{G}^{H}\mathbf{G} + \sigma_n^{2}\,\mathbf{I}\big)^{-1}
\mathbf{G}^{H}\,\mathbf{y}[s],
\label{eq:mmse}
\end{equation}
the standard regularized least-squares solution. Because $\mathbf{G}$ captures both the modal coupling and the delay spread, \eqref{eq:mmse} jointly suppresses IMI and ISI and coherently recombines the $L\times K$ modal-time diversity.

\begin{remark}
The receiver degenerates gracefully. When $\rho\to\infty$ (near-LoS), the cross-mode entries of $\mathbf{G}$ vanish, the MMSE solution factorizes into $L$ per-mode time-domain equalizers, and the 2D-Rake reduces to $L$ independent 1D (time-only) Rakes. When the delay spread vanishes ($K=1$), it reduces to a mode-domain MMSE receiver.
\end{remark}

\subsection{Complexity}
The receiver procedure is: DFT demultiplexing, construction of the symbol-level channel $\mathbf{G}$ from the CSI and the codes, and per-symbol combining \eqref{eq:finger}--\eqref{eq:mmse} with demapping. The computational cost separates into a \emph{per-coherence-block} part (channel estimation, $\mathbf{G}$ construction, and the Gram-matrix inverse, performed once per block because the channel is constant within a block) and a \emph{per-symbol} part (DFT, finger despreading, combining).

\begin{table}[t]
\centering\footnotesize\setlength{\tabcolsep}{2pt}
\caption{Computational complexity per symbol.}
\label{tab:complexity}
\begin{tabular}{@{}c c c c@{}}
\toprule
Receiver & Per-symbol cost & $L{=}4,K{=}4$ & $L{=}8,K{=}6$\\
\midrule
Mode-wise DFT detection & $O(L)$            & $4$             & $8$\\
Mode-domain ZF/MMSE & $O(L^{2})$ $^{\dagger}$ & $1.6\times10^{1}$ & $6.4\times10^{1}$\\
1D-Rake (time-only)   & $O(LKN_c)$        & $6.4\times10^{1}$ & $3.8\times10^{2}$\\
2D-Rake MRC           & $O(L^{2}KN_c)$    & $2.6\times10^{2}$ & $3.1\times10^{3}$\\
2D-Rake MMSE          & $O(L^{2}KN_c+L^{2})$ $^{\dagger}$ & $2.7\times10^{2}$ & $3.1\times10^{3}$\\
\bottomrule
\end{tabular}\\
\vspace{2pt}{\footnotesize$^{\dagger}$ plus the per-block construction of $\mathbf{G}$ at $O(L^{2}KWN_c)$ ($1.8\times10^{3}$ for $L{=}4,K{=}4,N_c{=}4$; $4.0\times10^{4}$ for $L{=}8,K{=}6,N_c{=}8$) and the $O(L^{3})$ Gram-matrix inversion. The $L{=}8$ column uses the shortest Walsh code with $N_c{=}8\ge L$.}
\end{table}

Table~\ref{tab:complexity} summarizes the cost. Per block the 2D-Rake MMSE pays $O(L^2KWN_c)$ for the $\mathbf{G}$ construction plus $O(L^3)$ for the inverse; per symbol it pays $O(L^2KN_c)$ for the fingers (all $L\times L$ mode pairs and all $K$ delays) plus $O(L^2)$ for the combining. The transmitter pays $O(LN_c)$ per symbol for the spreading. For reference, a conventional array-domain space-time MMSE equalizer must invert a Gram matrix of the full observation dimension, whereas the structured modal-time processing inverts only the $L\times L$ Gram matrix of the symbol-level model \eqref{eq:mmse} and exploits the $L\times K$ modal-delay structure explicitly.

\section{Diversity and Spectral-Efficiency Analysis}
\label{sec:diversity}

This section analyzes the reliability of the modal-time Rake with three distinct, and deliberately not interchangeable, notions: (i) the \emph{asymptotic diversity order} $d_{\mathrm{asym}}$, the high-SNR exponent of the error probability; (ii) an \emph{effective diversity metric} $D_{\mathrm{eff}}$ that characterizes finite-SNR behavior under unequal branch powers; and (iii) the \emph{combining dimensionality} $N_f=LK$ of the finger grid. In agreement with classical diversity theory \cite{goldsmith2005wireless}, the finger count is a combining dimensionality, not an unconditional diversity order: the diversity order is determined by the number of statistically independent, non-zero-power channel branches, equivalently by the rank of the channel covariance \cite{goldsmith2005wireless}.

\subsection{Channel fading models}
\label{sec:fading}
Two statistical fading models of the scattered taps lead, as this section shows, to different diversity behavior:
\begin{itemize}
\item \emph{Per-tap fading (physical model, used throughout Section~V).} Every mode pair of tap $k$ shares one Rayleigh coefficient, $h[\ell',\ell,k]=a_{\ell',\ell}\,g_k$, with $a_{\ell',\ell}$ the deterministic diffusion structure, so the fading dimension is $K$, not $LK$.
\item \emph{Independent-branch fading (idealized model).} Each coefficient carries its own independent Rayleigh factor, giving $LK$ independent fading branches, the model under which Theorem~1's $LK$-fold diversity applies, serving as the upper bound for the physical model.
\end{itemize}

\subsection{Diversity order under independent modal-delay branches}
\label{sec:divA}
Assume first the idealized case: the $LK$ finger coefficients $h[\ell',\ell,k]$ are statistically independent across the output-mode index $\ell'$ and the delay index $k$, and each has non-zero mean power. Each finger of the desired mode then carries an independently Rayleigh-faded copy of the symbol, and the MRC combiner \eqref{eq:mrc} is a maximal-ratio combiner of $LK$ i.i.d. branches. By the standard MRC result \cite{goldsmith2005wireless},
\begin{equation}
P_{e}\;\propto\;\big(E_{s}/N_{0}\big)^{-LK},
\label{eq:diversity}
\end{equation}
where $E_{s}/N_{0}$ is the symbol-energy-to-noise ratio, so that $d_{\mathrm{asym}}=LK$ for the MRC.

\begin{theorem}[Diversity under independent branches]
For the mode-time channel \eqref{eq:rx} whose $LK$ finger coefficients are mutually independent with non-zero average powers, both the MRC combiner \eqref{eq:mrc} and the joint MMSE combiner \eqref{eq:mmse} attain the asymptotic diversity order $d_{\mathrm{asym}}=LK$.
\end{theorem}
\begin{proof}[Sketch]
For MRC, the theorem's assumptions reduce the finger outputs of a target mode to $LK$ statistically independent Rayleigh observations of the symbol, so \eqref{eq:diversity} is the classical $LK$-branch result \cite{goldsmith2005wireless}. For the joint MMSE receiver, the modal-time observation \eqref{eq:obs} retains all $LK$ independent fading dimensions: the MRC weights form a feasible linear estimator, hence the conditional output signal-to-interference-plus-noise ratio of the MMSE solution is never smaller than that of the MRC in the same realization; for QPSK the symbol-error probability is a monotone decreasing function of that ratio, and averaging over the fading preserves the ordering, so $P_{e}^{\mathrm{MMSE}}\le P_{e}^{\mathrm{MRC}}\propto(E_s/N_0)^{-LK}$, i.e., the exponent is at least $LK$. Conversely, the observation carries no more than $LK$ independent fading dimensions, so no estimator can exceed the exponent $LK$. \end{proof}

\subsection{Effective diversity metric under unequal branch powers}
\label{sec:divB}
Independence alone does not guarantee that all branches are useful at \emph{finite} SNR. With a finite mode coherence $\rho$, the modal branches of a given delay tap are unequal-power: the self-mode branch carries unit power and each of the $L-1$ cross-mode branches carries the fraction
\begin{equation}
\gamma \;=\; \frac{1}{(L-1)\rho}
\label{eq:gamma}
\end{equation}
of it, which follows from the total cross-mode power definition \eqref{eq:coherence}. For MRC over $N_f$ unequal-power but independent Rayleigh branches with powers $p_i$, the finite-SNR behavior is governed by the quantity $(\sum_i p_i)^{2}/\sum_i p_i^{2}$ \cite{goldsmith2005wireless}. With $K$ self-mode branches of power $P$ and $K(L-1)$ cross-mode branches of power $\gamma P$,
\begin{equation}
D_{\mathrm{eff}} \;=\; K\,\frac{\big(1+(L-1)\gamma\big)^{2}}{1+(L-1)\gamma^{2}}, \qquad \gamma=\big((L-1)\rho\big)^{-1}.
\label{eq:deff}
\end{equation}
Equation~\eqref{eq:deff} is an \emph{effective} diversity metric: it measures how much of the full combining dimensionality remains usable when the cross-mode branches are weak, which is the relevant figure at finite SNR. It is \emph{not} the asymptotic diversity order: whenever the $LK$ branches are independent with non-zero power, the high-SNR exponent remains $LK$ by Theorem~1, and a weak branch merely delays the onset of the asymptotic slope. Its limits are informative: $\rho\to\infty$ ($\gamma\to0$) gives $D_{\mathrm{eff}}\to K$, while $\gamma\to1$, which corresponds to $\rho=1/(L-1)$ and equally powered branches, gives $D_{\mathrm{eff}}\to KL$. At the operating point of Section~V ($\rho\approx4.4$, $L=4$, $K=4$) it evaluates to $\approx5.9$ of the maximum $16$. Since \eqref{eq:deff} presumes independent modal branches, in the per-tap fading model of Section~\ref{sec:fading}, where the branches are correlated through the shared tap gain, it overstates the modal contribution and the covariance-rank analysis of Section~\ref{sec:divC} governs instead.

\subsection{Diversity under correlated modal diffusion}
\label{sec:divC}
The Gaussian mode-diffusion model \eqref{eq:diffusion} couples the modes through a common scattering mechanism, so the finger coefficients $h[\ell',\ell,k]$ and $h[\ell'',\ell,k]$ sharing the same input mode and delay need not be independent. Correlated branches reduce the diversity, and the upper bound is then set by the rank of the channel covariance rather than by the finger count. For a target mode $\ell$, define the modal branch vector of the $k$-th resolvable delay tap as
\begin{equation}
\mathbf{h}_{k,\ell} \;=\; \big[h_{1,\ell,k},\;h_{2,\ell,k},\;\dots,\;h_{L,\ell,k}\big]^{T}\;\in\;\mathbb{C}^{L},
\label{eq:branchvec}
\end{equation}
i.e., the $L$ branches that the 2D-Rake actually combines to recover mode $\ell$ at tap $k$, and its modal covariance matrix as
\begin{equation}
\mathbf{R}_{k,\ell} \;=\; \mathbb{E}\!\big[\mathbf{h}_{k,\ell}\mathbf{h}_{k,\ell}^{H}\big],
\label{eq:cov}
\end{equation}
with eigenvalues $\lambda_1,\dots,\lambda_{L}$. The rank of $\mathbf{R}_{k,\ell}$ characterizes how many statistically independent modal fading dimensions tap $k$ offers for recovering the target mode, and the effective rank
\begin{equation}
r_{\mathrm{eff},k,\ell} \;=\; \frac{\big(\sum_i \lambda_i\big)^{2}}{\sum_i \lambda_i^{2}}\;\le\;L
\label{eq:reff}
\end{equation}
measures how many of them remain usable under unequal powers. The following bound, standard for covariance-limited combining \cite{goldsmith2005wireless}, then applies.

\begin{theorem}[Rank-limited diversity]
For a target mode $\ell$ of the mode-time channel \eqref{eq:rx}, assuming statistically independent fading across the resolvable delay taps, the asymptotic diversity order of any linear combiner (including \eqref{eq:mrc} and \eqref{eq:mmse}) is upper-bounded by
\begin{equation}
d_{\mathrm{asym},\ell} \;\le\; \sum_{k=1}^{K} \mathrm{rank}\big(\mathbf{R}_{k,\ell}\big) \;\le\; LK,
\label{eq:rankbound}
\end{equation}
with equality $d_{\mathrm{asym},\ell}=LK$ when every tap contributes $L$ statistically independent, non-zero-power modal branches, which is exactly the independent-branch case of Theorem~1. When the taps are correlated, the bound generalizes to $d_{\mathrm{asym},\ell}\le\mathrm{rank}(\mathbf{R}_\ell)$ with the full modal-delay covariance $\mathbf{R}_\ell=\mathbb{E}[\mathbf{h}_\ell\mathbf{h}_\ell^{H}]$, which reduces to \eqref{eq:rankbound} for independent taps. Under full-rank but \emph{correlated} modal diffusion, the finite-SNR behavior is governed by $K\cdot r_{\mathrm{eff}}$, with $r_{\mathrm{eff}}$ the target-mode effective rank \eqref{eq:reff} averaged over taps, and \eqref{eq:deff} remains the unequal-power metric.
\end{theorem}

\begin{remark}[A cautionary example]
If the tap coupling factorizes as $h[\ell',\ell,k]=a_{\ell',\ell}\,g_k$, then $\mathbf{h}_{k,\ell}=g_k\,[a_{1,\ell},\dots,a_{L,\ell}]^{T}$, the covariance $\mathbf{R}_{k,\ell}$ has rank one, and tap $k$ offers a single independent fading dimension: the modal dimension adds energy but no independent diversity despite the $L$-fold finger count. Counting fingers without inspecting the covariance overestimates the diversity.
\end{remark}

\subsection{When is IMI diversity, and when is it interference?}
\label{sec:divD}
The paper-level message, \emph{multipath-induced modal leakage can be recovered as diversity}, holds only if four ingredients are present; if any of them fails, the leaked energy turns back into interference: sufficient cross-modal energy, separability by the colored codes, sufficiently low modal correlation (rank), and adequate CSI.
Specifically: (i) if the leakage power $\gamma$ is too small, the harvested energy is negligible; (ii) if the codes lose zero-lag orthogonality (e.g., fractional timing offsets, Section~\ref{sec:timing}), the modes cannot be separated; (iii) if the covariance rank collapses (the example above), the leaked branches add energy but not diversity; and (iv) if the CSI is poor, the combining weights are wrong. Section~\ref{sec:rank} quantifies this interplay and the operating window in which IMI converts into diversity.

\subsection{Outage probability}
Under the idealized independent-branch model of Theorem~1, with $N_f$ i.i.d. Rayleigh branches of mean SNR $\bar{\gamma}$ each, the MRC output SNR is $\chi^2$-distributed with $2N_f$ degrees of freedom, and the outage probability at threshold $\gamma_{\mathrm{th}}$ is
\begin{equation}
P_{\mathrm{out}} \;=\; 1 - e^{-\gamma_{\mathrm{th}}/\bar{\gamma}}
\sum_{k=0}^{N_f-1}\frac{(\gamma_{\mathrm{th}}/\bar{\gamma})^{k}}{k!}
\;\sim\;
\frac{(\gamma_{\mathrm{th}}/\bar{\gamma})^{N_f}}{N_f!},
\label{eq:outage}
\end{equation}
valid for $\bar{\gamma}\gg\gamma_{\mathrm{th}}$, i.e., $P_{\mathrm{out}}$ decays as $\bar{\gamma}^{-N_f}$, the same exponent as \eqref{eq:diversity}. For a $10\%$ outage target, the required mean branch SNR is $\bar{\gamma}=\gamma_{\mathrm{th}}/q_{N_f}$ ($q_4\approx1.75$, $q_{16}\approx11.1$); the 2D-Rake thus attains $\bar{\gamma}\approx0.09\,\gamma_{\mathrm{th}}$, versus $0.57\,\gamma_{\mathrm{th}}$ for the time-only Rake, a saving of about $\mathbf{8}$~dB.

\subsection{Spectral-efficiency trade-off}
\label{sec:divE}
We define the normalization explicitly. Let the chip duration be $T_c$ and the transmit power $P_t$; both systems occupy the same bandwidth $1/T_c$, the same frame duration $T_{\mathrm{frame}}=N_{\mathrm{slot}}T_c$, and the same transmit power. The unspread system transmits one QPSK symbol per mode per chip slot, so its information rate and energy per bit are
\begin{equation}
R_{b}^{\mathrm{(u)}} = \frac{2L\,N_{\mathrm{slot}}}{T_{\mathrm{frame}}},
\qquad
E_{b}^{\mathrm{(u)}} = \frac{P_t}{R_{b}^{\mathrm{(u)}}},
\label{eq:se1}
\end{equation}
where $N_{\mathrm{slot}}$ is the number of chip slots in the frame. The colored system spreads each symbol over $N_c$ chips and appends $N_g=K-1$ guard chips, so within the same frame it transmits only $N_{\mathrm{slot}}/(N_c+N_g)$ symbols per mode: its information rate drops by the factor
\begin{equation}
\eta \;=\; \frac{1}{N_c+N_g},
\label{eq:se}
\end{equation}
and, at the same transmit power, its energy per bit rises by the reciprocal factor, $E_{b}^{\mathrm{(c)}}=(N_c+N_g)E_{b}^{\mathrm{(u)}}$. At a common $E_b/N_0$, this normalization therefore attributes the colored system a $10\log_{10}(N_c+N_g)\approx8.5$~dB penalty for $N_c=4,N_g=3$ (the curve shift in Fig.~\ref{fig:se}); the shift is an artifact of the fixed-frame comparison, not a claim about other normalizations. In spectral-efficiency (SE) terms, the unspread QPSK system carries $\mathrm{SE}^{(\mathrm{u})}=2L$ bit/s/Hz, the colored one $\mathrm{SE}^{(\mathrm{c})}=2L\eta$ bit/s/Hz, and with an $M$-ary constellation $\mathrm{SE}^{(\mathrm{c})}=L\eta\log_2 M$ bit/s/Hz. The rate loss is thus the price of the diversity, not of bandwidth, and it is partly recoverable: because the 2D-Rake lowers the $E_b/N_0$ required at a target BER, higher-order constellations become affordable, which is not an option for the interference-floor-limited conventional detector. Section~\ref{sec:setradeoff} reports the quantitative trade-off between the code length, the required SNR, and the achievable SE at a fixed target BER.

\subsection{Adaptive coloring and conditioning}
Coloring pays only when the channel is rich: in a LoS-dominant channel ($\rho$ large) the modal diversity vanishes and the spreading overhead \eqref{eq:se} buys little, whereas in rich multipath it converts IMI into diversity. The transmitter can therefore key the coloring on the estimated mode coherence, $\text{coloring}=\text{on}$ if $\hat{\rho}<\rho_{\mathrm{th}}$ and off otherwise; Fig.~\ref{fig:nclos}(b) quantifies this dependence (for a LoS fraction above $\approx0.8$ the conventional detector comes within about $15$ times of the 2D-Rake, and switching off saves the full penalty $\eta$), and the optimal threshold is a one-dimensional search left to the implementation.

For the conditioning of the MMSE combiner \eqref{eq:mmse}, the condition number $\kappa$ of the Gram matrix $\mathbf{G}^{H}\mathbf{G}$ stays between $2.5$ and $2.9$ over the \emph{tested} angular spreads of Section~V, so the regularization $\sigma_n^2\mathbf{I}$ suffices there; we do not claim this beyond the tested range.

\section{Numerical Results}
\label{sec:results}

\subsection{Simulation setup}
Table~\ref{tab:params} lists the simulation parameters. All numerical validation, simulation, and plotting were performed in MATLAB. The OAM generation follows the free-space UCA field model of \eqref{eq:arrayfactor} (isotropic elements, $\lambda/2$ spacing); the communication simulation uses the mode-time channel \eqref{eq:rx} with the Gaussian mode-diffusion kernel \eqref{eq:diffusion} and a Rician LoS/non-LoS split. Unless stated otherwise, perfect CSI is assumed at the receiver.

\emph{Reproducibility.} Each experiment reports the bit count per point (e.g., $3.2\times10^{4}$ in Fig.~\ref{fig:ber}, $4.8\times10^{4}$ in Section~V-F), the number of channel realizations, and the seed policy; BERs of the form ``$<x$'' mean zero observed errors over the stated bit count ($\mathrm{BER}<1/N_{\mathrm{bits}}$). The complete MATLAB simulation and figure-generation code is available from the authors.

\begin{table}[t]
\centering
\caption{Simulation parameters.}
\label{tab:params}
\begin{tabular}{@{}l l@{}}
\toprule
Parameter & Value\\
\midrule
Carrier frequency $f_c$ & 28~GHz ($\lambda\approx10.7$~mm)\\
UCA elements $N$ & 8, radius $R=N\lambda/(4\pi)\approx6.8$~mm\\
Active modes $L$ & 4: $\ell\in\{-2,-1,+1,+2\}$\\
Multipath taps $K$ & 4 (exponential PDP)\\
Spreading code & Walsh--Hadamard, $N_c=4$, guard $N_g=3$\\
Modulation & QPSK\\
Angular spread $\Delta\theta$ & $35^{\circ}$ (operating point)\\
LoS power fraction & 0.3; coupling strength 1.0\\
\bottomrule
\end{tabular}
\end{table}

\subsection{OAM beam generation and coupling-law validation}
\begin{figure}[t]
\centering
\subfigure[Mode $\ell=-2$.]{\includegraphics[width=0.47\columnwidth]{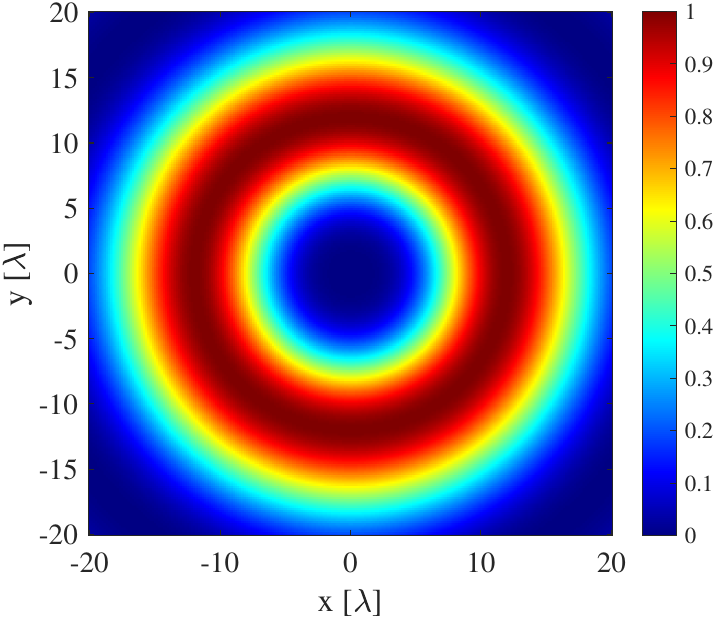}}
\subfigure[Mode $\ell=+1$.]{\includegraphics[width=0.47\columnwidth]{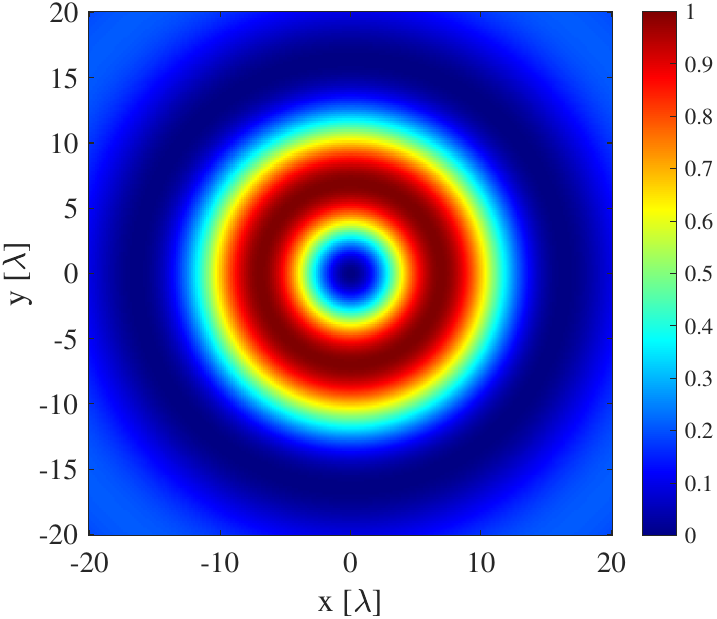}}
\caption{Simulated intensity of the free-space OAM beams for $\ell=-2$ and $\ell=+1$.}
\label{fig:beams}
\end{figure}

\begin{figure}[t]
\centering
\subfigure[Per-ray coupling law \eqref{eq:bessel_law}.]{\includegraphics[width=0.47\columnwidth]{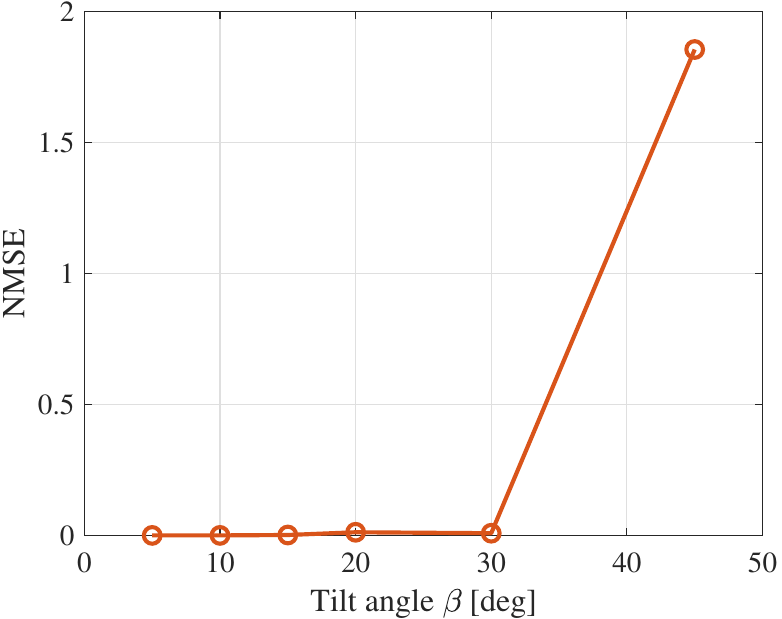}}
\subfigure[Mode-diffusion kernel \eqref{eq:diffusion}.]{\includegraphics[width=0.47\columnwidth]{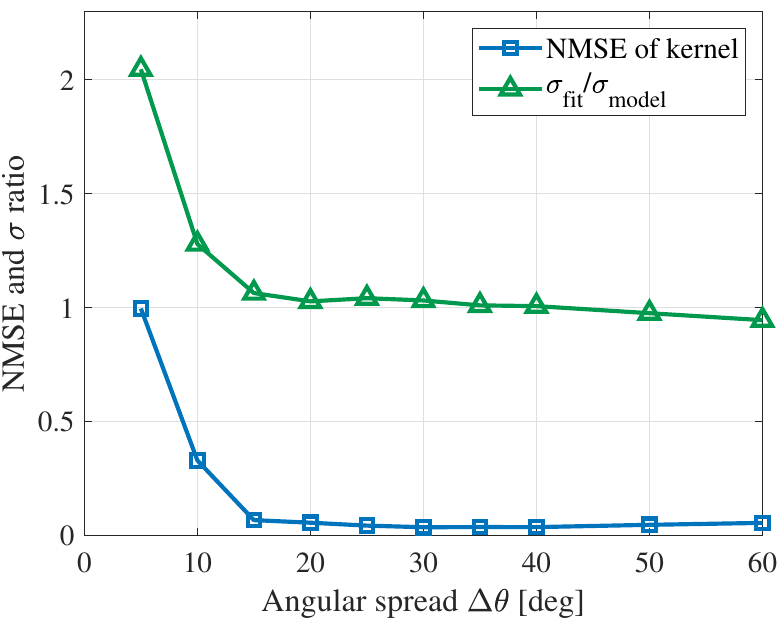}}
\caption{Validation of the mode-coupling channel model.}
\label{fig:valnmse}
\end{figure}

Figure~\ref{fig:beams} shows the normalized transverse intensity of the free-space beams radiated by a $16$-element UCA at $z=30\lambda$ under the isotropic-element field model of \eqref{eq:arrayfactor}: the phase excitation of \eqref{eq:txphase} produces the expected intensity rings for both modes. The $16$-element array serves only this independent validation; the communication simulations use the $8$-element UCA of Table~\ref{tab:params}. The coupling law \eqref{eq:bessel_law} is validated quantitatively: the normalized spiral-spectrum coupling of a beam tilted by $\beta=30^{\circ}$, computed by full free-space field propagation, deviates from the Bessel prediction by less than $0.13$ per entry over $d=-4,\dots,4$. Figure~\ref{fig:valnmse}(a) quantifies the agreement over the tilt-angle range: the NMSE of \eqref{eq:bessel_law} is below $1.2\%$ for $\beta\le30^{\circ}$ and grows sharply beyond (NMSE$=1.85$ at $\beta=45^{\circ}$), delineating the validity region of the far-field law; the residual small-angle asymmetry is attributable to the finite-distance Fresnel terms that \eqref{eq:couplinglaw} neglects. The $30^{\circ}$ boundary concerns a single ray of fixed tilt $\beta$; the operating $\Delta\theta=35^{\circ}$ is the RMS width of the tilt distribution, over which the ensemble kernel \eqref{eq:diffusion} is validated (Fig.~\ref{fig:valnmse}(b)).

The Gaussian mode-diffusion kernel \eqref{eq:diffusion} approximates the exact angular average of \eqref{eq:bessel_law}. Figure~\ref{fig:valnmse}(b) compares it to the Monte-Carlo average of the normalized Bessel amplitudes: for $\Delta\theta\ge15^{\circ}$ the NMSE is $3.5\%$--$5.5\%$, and the fitted bandwidth matches $\sigma_\ell=kR\sin(\Delta\theta)/\sqrt2$ within $1.4\%$ at the operating point ($\hat{\sigma}_\ell=1.645$ vs.\ $1.622$). At very small spreads the relative NMSE is dominated by near-zero entries, and both predict essentially diagonal channels. We therefore adopt \eqref{eq:diffusion} as a validated tractable model over the operating range, noting that real scattering statistics are generally non-Gaussian \cite{paterson2005turbulence, rodenburg2012turbulence}.

\subsection{Mode coherence}
\begin{figure}[t]
\centering
\includegraphics[width=0.82\columnwidth]{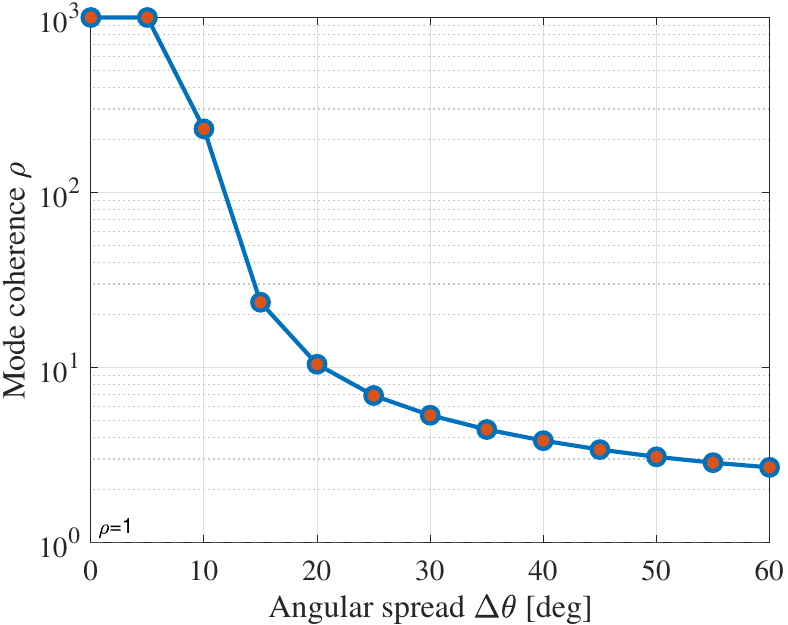}
\caption{Mode coherence $\rho$ versus the RMS angular spread $\Delta\theta$.}
\label{fig:coherence}
\end{figure}

Figure~\ref{fig:coherence} shows the mode coherence $\rho$ \eqref{eq:coherence} as a function of the angular spread, averaged over channel realizations: $\rho$ falls from effectively infinite at $\Delta\theta=0$ to $\approx2.7$ at $\Delta\theta=60^{\circ}$, with $\rho\approx4.4$ at the operating point $\Delta\theta=35^{\circ}$. The monotonic decay quantifies the ``degradation but not destruction'' of the modal orthogonality; combined with the rank analysis of Section~\ref{sec:rank}, it delineates the operating window of the receiver (the energy-to-rank condition of Section~\ref{sec:divD}).

\subsection{BER performance}
\begin{figure}[t]
\centering
\includegraphics[width=0.82\columnwidth]{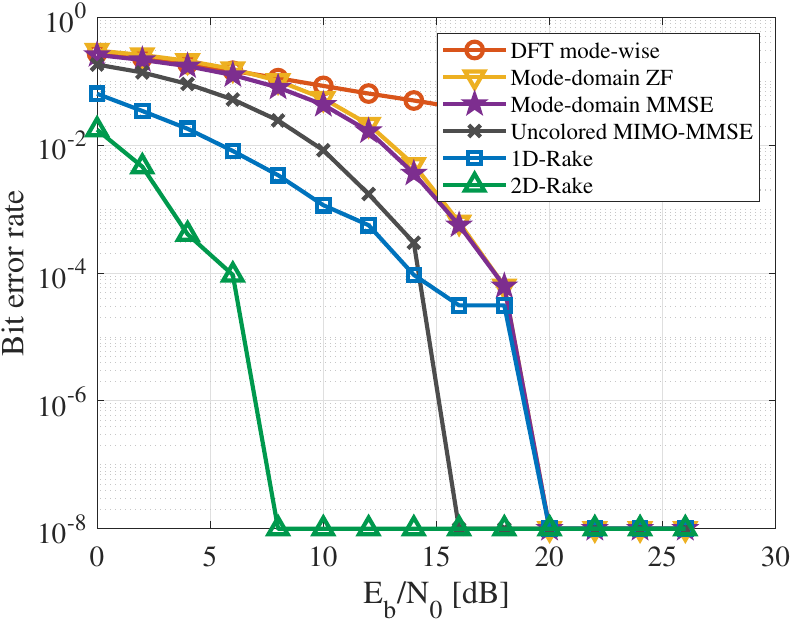}
\caption{Bit-error rate versus $E_b/N_0$ for the receivers of Table~\ref{tab:baselines}.}
\label{fig:ber}
\end{figure}

\begin{table}[t]
\centering
\caption{Receiver baseline hierarchy.}
\label{tab:baselines}
\footnotesize\setlength{\tabcolsep}{1pt}
\begin{tabular}{@{}l c c l@{}}
\toprule
Receiver & Cross-mode CSI & Delay comb. & Role\\
\midrule
DFT mode-wise & no & no & traditional OAM-MDM\\
Mode-domain ZF & yes & no & delay-collapsed MIMO\\
Mode-domain MMSE & yes & no & delay-collapsed MIMO\\
Uncolored MMSE & yes & yes & full space-time MIMO\\
1D-Rake & mode-matched & yes & time-only ablation\\
2D-Rake (proposed) & yes & yes & proposed\\
\bottomrule
\end{tabular}
\end{table}

Figure~\ref{fig:ber} and Table~\ref{tab:ber} give the core comparison over the baseline hierarchy of Table~\ref{tab:baselines} at $\Delta\theta=35^{\circ}$, $L=K=N_c=4$ with perfect CSI, where markers at the $10^{-8}$ axis floor denote zero observed errors over $3.2\times10^{4}$ bits per point. (The mode-domain ZF/MMSE detectors operate on the delay-collapsed coupling $\sum_k\mathbf{H}_k$ of the \emph{unspread} system; since the DFT across the UCA is unitary, they are equivalent to narrowband array-domain MIMO ZF/MMSE detectors. The uncolored modal-time MMSE baseline retains the delay dimension but not the temporal signatures.) The mode-wise DFT detector saturates at a floor of $\approx1.8\times10^{-2}$, because it discards the leaked modal energy. The mode-domain ZF/MMSE detectors recover it and reach the finite-sample resolution ($<3.1\times10^{-5}$ at $3.2\times10^{4}$ bits per point) beyond 18--20~dB, but they trail the 2D-Rake by $6$--$12$~dB. The uncolored modal-time MMSE equalizer, which retains the delay dimension but not the temporal signatures, recovers faster ($1.8\times10^{-1}$ at $0$~dB, $1.7\times10^{-3}$ at $12$~dB), yet the 2D-Rake stays ahead at every SNR, by up to two orders of magnitude at $4$--$8$~dB. The mode-matched 1D-Rake, which keeps only the temporal diversity, reaches $6.4\times10^{-2}$ at $0$~dB and $3.4\times10^{-3}$ at $8$~dB, so it needs $1$--$6$~dB more SNR than the 2D-Rake at the target BERs of Fig.~\ref{fig:snrgain}. The 2D-Rake itself yields $\mathbf{1.8\times10^{-2}}$ at $0$~dB, $4.1\times10^{-4}$ at $4$~dB, and zero observed errors beyond $8$~dB. The hierarchy shows that the structured modal-time processing beats the delay-collapsed and the uncolored full MIMO equalizers alike, and that time-only diversity is insufficient on this channel.

\begin{table}[t]
\centering\footnotesize\setlength{\tabcolsep}{1pt}
\caption{BER comparison at $\Delta\theta=35^{\circ}$, $L=4$, $K=4$, $N_c=4$, in dB (perfect CSI; $3.2\times10^{4}$ bits per point; U-MMSE denotes the uncolored modal-time MMSE equalizer).}
\label{tab:ber}
\begin{tabular}{@{}c c c c c c c@{}}
\toprule
$E_b/N_0$ & DFT & ZF & MMSE & U-MMSE & 1D-Rake & \textbf{2D-Rake}\\
\midrule
0  & $2.7\mathrm{e}{-1}$ & $3.0\mathrm{e}{-1}$ & $2.6\mathrm{e}{-1}$ & $1.8\mathrm{e}{-1}$ & $6.4\mathrm{e}{-2}$ & $\mathbf{1.8\mathrm{e}{-2}}$\\
4  & $1.9\mathrm{e}{-1}$ & $2.1\mathrm{e}{-1}$ & $1.7\mathrm{e}{-1}$ & $9.1\mathrm{e}{-2}$ & $1.8\mathrm{e}{-2}$ & $\mathbf{4.1\mathrm{e}{-4}}$\\
8  & $1.1\mathrm{e}{-1}$ & $1.0\mathrm{e}{-1}$ & $8.0\mathrm{e}{-2}$ & $2.5\mathrm{e}{-2}$ & $3.4\mathrm{e}{-3}$ & $\mathbf{<3.1\mathrm{e}{-5}}$\\
12 & $6.5\mathrm{e}{-2}$  & $2.1\mathrm{e}{-2}$  & $1.7\mathrm{e}{-2}$  & $1.7\mathrm{e}{-3}$  & $5.6\mathrm{e}{-4}$ & $\mathbf{<3.1\mathrm{e}{-5}}$\\
16 & $3.9\mathrm{e}{-2}$  & $6.3\mathrm{e}{-4}$  & $5.6\mathrm{e}{-4}$  & $<3.1\mathrm{e}{-5}$ & $3.1\mathrm{e}{-5}$ & $\mathbf{<3.1\mathrm{e}{-5}}$\\
20 & $2.5\mathrm{e}{-2}$  & $<3.1\mathrm{e}{-5}$ & $<3.1\mathrm{e}{-5}$ & $<3.1\mathrm{e}{-5}$ & $<3.1\mathrm{e}{-5}$ & $\mathbf{<3.1\mathrm{e}{-5}}$\\
24 & $2.0\mathrm{e}{-2}$  & $<3.1\mathrm{e}{-5}$ & $<3.1\mathrm{e}{-5}$ & $<3.1\mathrm{e}{-5}$ & $<3.1\mathrm{e}{-5}$ & $\mathbf{<3.1\mathrm{e}{-5}}$\\
\bottomrule
\end{tabular}
\end{table}

The mode-matched 1D-Rake warrants a comment, since trailing even the naive mode-wise detector at low SNR is counter-intuitive. The delay-shifted cross-correlations $\mathbf{c}_\ell^{H}\mathbf{c}_{\ell'}(\tau)$, $\tau\neq0$, of the Walsh--Hadamard codes are non-zero and large, so a Rake that combines over the delay taps re-injects delay-shifted inter-code interference into every finger, whereas zero-lag orthogonality protects only same-delay separation. This is what limits the time-only 1D-Rake and why the joint MMSE combiner is needed rather than optional. The MRC combiner of \eqref{eq:mrc} was also simulated: the 2D modal-time MRC reaches $2.7\times10^{-2}$, $4.0\times10^{-3}$ and $9.4\times10^{-5}$ at $0$, $4$ and $8$~dB, i.e., $1.5$, ten and three times worse than the joint MMSE, and the time-only MRC is a further $1.5$--$2.7$ times worse; the gap is the delay-shifted residue that only the joint equalizer cancels. This ablation shows that delay diversity alone is insufficient, and that \eqref{eq:mmse} is necessary.

\subsection{SNR gain at target BER}
\begin{figure}[t]
\centering
\includegraphics[width=0.82\columnwidth]{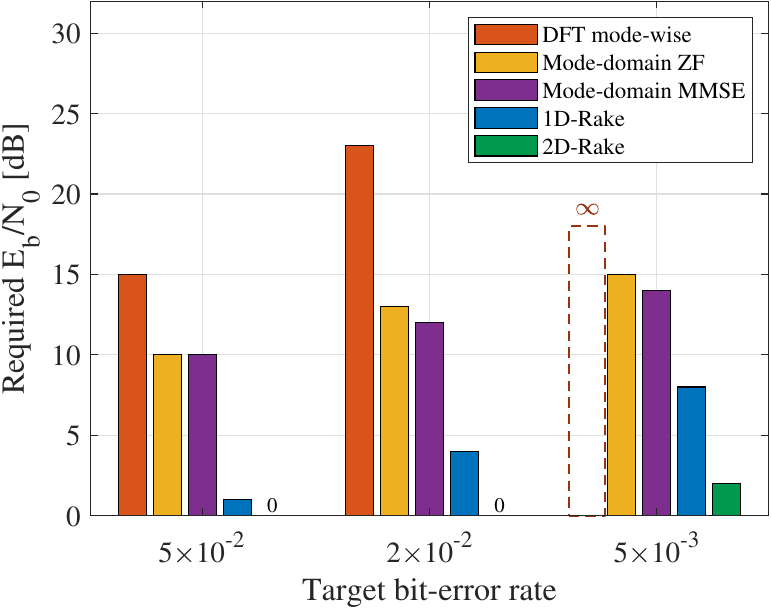}
\caption{Required $E_b/N_0$ versus the target BER at $\Delta\theta=35^{\circ}$.}
\label{fig:snrgain}
\end{figure}

Fig.~\ref{fig:snrgain} converts the curves of Fig.~\ref{fig:ber} into required-$E_b/N_0$ at target BERs: a label ``0' marks a receiver that attains the target already at $0$~dB, while the $\infty$ symbol drawn over a dashed placeholder bar marks the mode-wise DFT detector at the $5\times10^{-3}$ target, which no finite SNR attains because its interference floor lies above it. At $5\times10^{-2}$ the 2D-Rake needs $\approx0$~dB, against $\approx10$~dB for the mode-domain MMSE/ZF detectors and $15$~dB for the mode-wise detector, which is a saving of $\mathbf{10}$--$\mathbf{15}$~dB; at $2\times10^{-2}$ the savings grow to about $\mathbf{12}$~dB for the MMSE detector, $\mathbf{13}$~dB for the ZF detector and $\mathbf{23}$~dB for the mode-wise detector, while the 1D-Rake still needs $4$~dB; at $5\times10^{-3}$ the mode-wise detector cannot reach the target at all, because its floor lies above it, whereas the 1D-Rake attains it at $8$~dB and the 2D-Rake at $\approx2$~dB.

\subsection{Diversity order under block fading}
\begin{figure}[t]
\centering
\subfigure[Versus taps $K$, with $L=4$.]{\includegraphics[width=0.47\columnwidth]{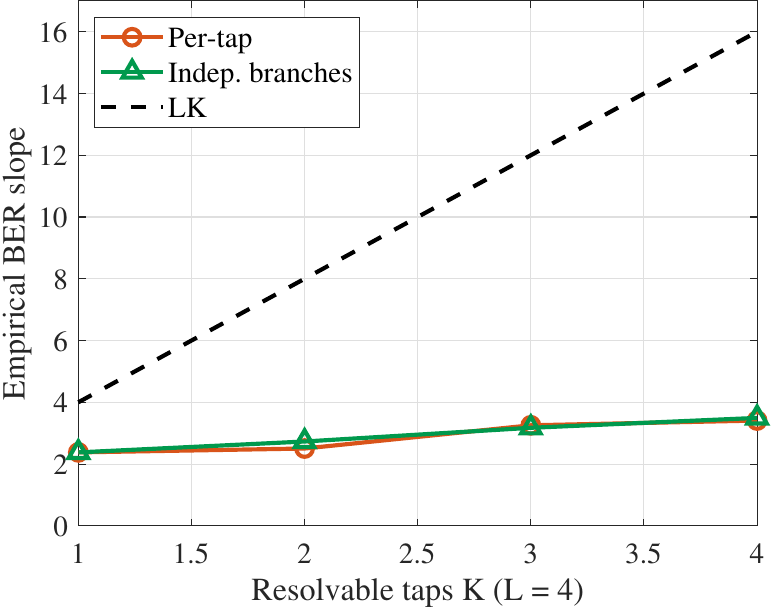}}
\subfigure[Versus modes $L$, with $K=2$.]{\includegraphics[width=0.47\columnwidth]{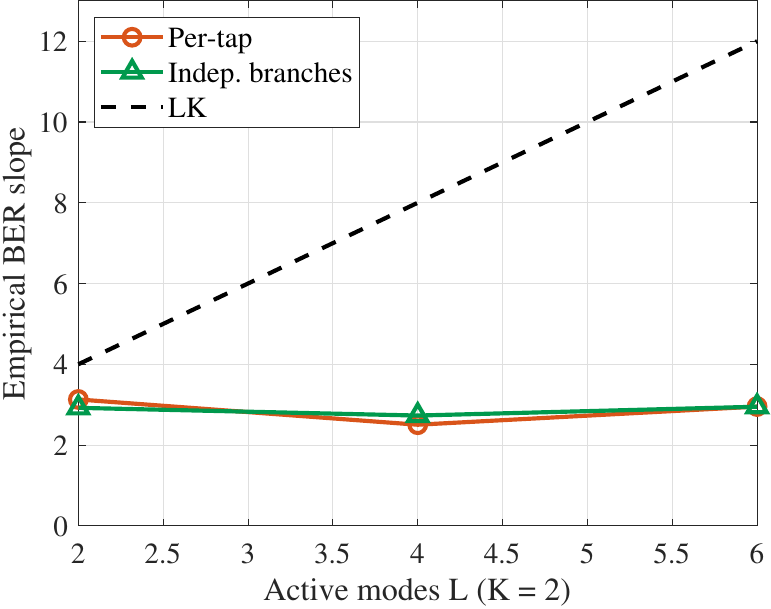}}
\caption{Empirical BER slope versus the tap count $K$ (a) and the mode count $L$ (b); the curves are not asymptotic diversity-order estimates.}
\label{fig:divK}
\end{figure}

Averaging the BER over independent channel realizations (block fading, $4.8\times10^{4}$ bits per point), the mode-wise detector remains at a high floor ($0.32$ at $0$~dB, falling only to $0.14$ at $24$~dB), the 1D-Rake falls from $5.0\times10^{-2}$ to about $7\times10^{-3}$ and then flattens, and the 2D-Rake drops below the finite-sample resolution already at $6$~dB, with an empirical exponent of $\approx3.4$ over the observable window. We report this as an \emph{empirical reliability slope}, not as a verification of a specific diversity order: the finite-sample resolution truncates the high-SNR region and, more fundamentally, the adopted per-tap channel is not an independent-branch channel. In that model (Section~\ref{sec:fading}) all mode pairs of a tap share one fading coefficient, so the modal dimension supplies \emph{energy} rather than independent diversity, and the asymptotic order is set by the number of independent tap gains ($K$), not by $LK$; the $LK$ order of Theorem~1 applies to the independent-branch channel, whose slope lies beyond the present resolution. Figure~\ref{fig:divK} confirms the scaling: the measured slope grows from $2.4$ ($K=1$) to $3.4$ ($K=4$), and Section~\ref{sec:rank} measures the target-mode effective rank directly (at most $\approx2.9$ for the independent-branch channel, exactly $1$ for the per-tap channel).

\subsection{Parameter analysis}
\begin{figure}[t]
\centering
\includegraphics[width=0.82\columnwidth]{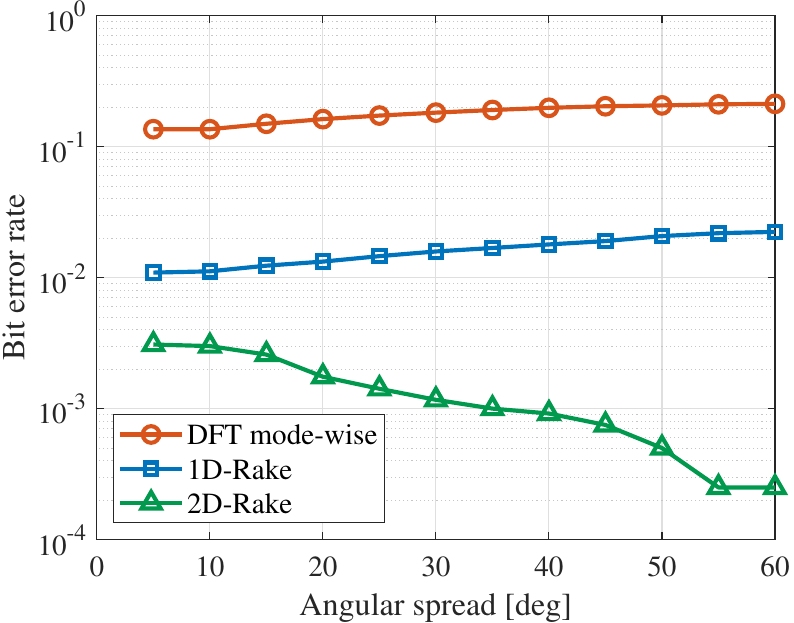}
\caption{BER versus the RMS angular spread at $E_b/N_0=4$~dB.}
\label{fig:dtheta}
\end{figure}

Figure~\ref{fig:dtheta} shows the BER as a function of the angular spread. Both baselines degrade monotonically as $\Delta\theta$ grows ($0.14\to0.21$ for the mode-wise detector and $1.1\times10^{-2}\to2.2\times10^{-2}$ for the 1D-Rake). The 2D-Rake exhibits the opposite behavior within the tested range: its BER \emph{falls} from $3.1\times10^{-3}$ at $5^{\circ}$ to $2.5\times10^{-4}$ at $60^{\circ}$. The improvement does not imply that stronger multipath is intrinsically beneficial: under the power-normalized channel model, larger spreads redistribute energy from the self-mode branch into separable cross-mode branches. We state this conditionally: the benefit holds under the adopted mode-diffusion model and the tested range, so it is not a universal law; Section~\ref{sec:rank} gives the companion condition (sufficient modal rank) for the harvested energy to become reliability.

\begin{figure}[t]
\centering
\subfigure[Versus taps $K$.]{\includegraphics[width=0.47\columnwidth]{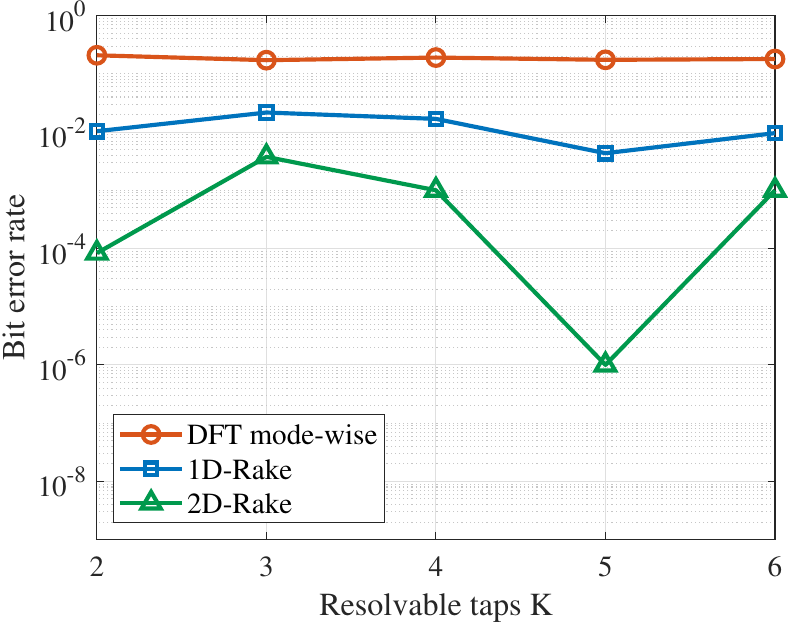}}
\subfigure[Versus modes $L$.]{\includegraphics[width=0.47\columnwidth]{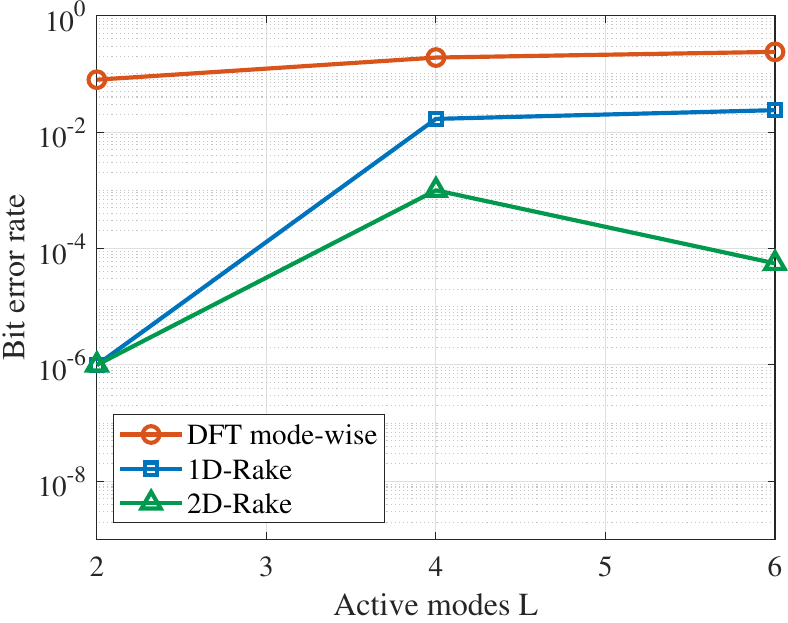}}
\caption{BER versus the number of resolvable taps $K$ (a) and the number of active modes $L$ (b), at $E_b/N_0=4$~dB.}
\label{fig:KL}
\end{figure}

\begin{figure}[t]
\centering
\subfigure[Versus code length $N_c$.]{\includegraphics[width=0.47\columnwidth]{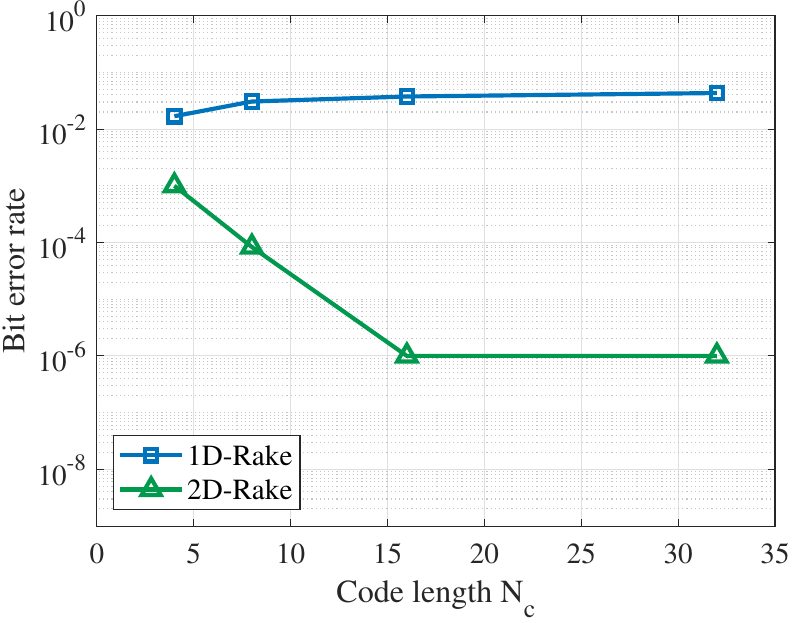}}
\subfigure[Versus LoS power fraction.]{\includegraphics[width=0.47\columnwidth]{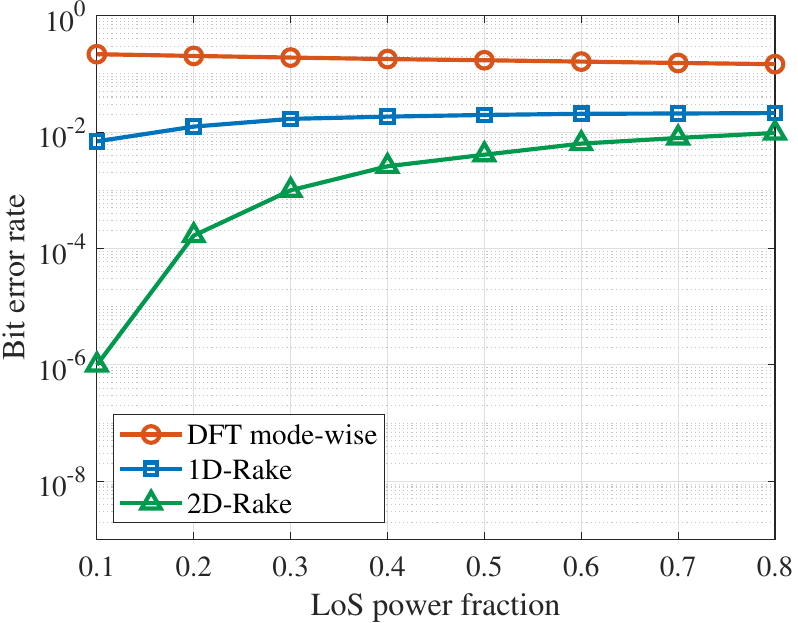}}
\caption{BER versus the code length $N_c$ (a) and the LoS power fraction (b), at $E_b/N_0=4$~dB.}
\label{fig:nclos}
\end{figure}

Figure~\ref{fig:KL}(a) varies the tap count: the 2D-Rake stays below $4\times10^{-3}$ for all $K$, while the mode-wise detector fluctuates around $0.2$ and the 1D-Rake between $4.3\times10^{-3}$ ($K=5$) and $2.2\times10^{-2}$ ($K=3$). Figure~\ref{fig:KL}(b) varies the mode count: the mode-wise detector degrades from $0.08$ ($L=2$) to $0.24$ ($L=6$) because more modes inject more IMI, whereas the 2D-Rake holds at $\approx10^{-3}$ or below, since the joint MMSE suppresses the IMI of all modes simultaneously. Figure~\ref{fig:nclos} extends the analysis to the code length and the LoS fraction.

Figure~\ref{fig:nclos}(a): longer codes lower the 2D-Rake BER ($10^{-3}$ at $N_c=4$ to below $10^{-4}$ at $N_c=16$), at the cost of the larger spreading penalty \eqref{eq:se}. Figure~\ref{fig:nclos}(b): as the channel becomes LoS-dominant (LoS fraction $0.1\to0.8$), the mode-wise detector improves ($0.22\to0.15$) while the 2D-Rake degrades mildly, from below the finite-sample resolution at LoS fraction $0.1$ to $\approx10^{-2}$ at $0.8$, approaching the time-only limit $\rho\to\infty$ of \eqref{eq:deff}; the receiver thus maintains a substantial advantage over the tested Rician power range. Finally, growing the element count $N$ at fixed $\lambda/2$ spacing degrades the baselines ($0.17\to0.23$ for $N=6\to16$) while the 2D-Rake improves from $1.4\times10^{-3}$ to $1.7\times10^{-4}$; this is a fixed-spacing aperture-scaling study, and element-level aliasing lies outside the mode-domain model.

\subsection{When does IMI become diversity: energy versus rank}
\label{sec:rank}
\begin{figure*}[t]
\centering
\subfigure[Mode coherence $\rho$.]{\includegraphics[width=0.32\textwidth]{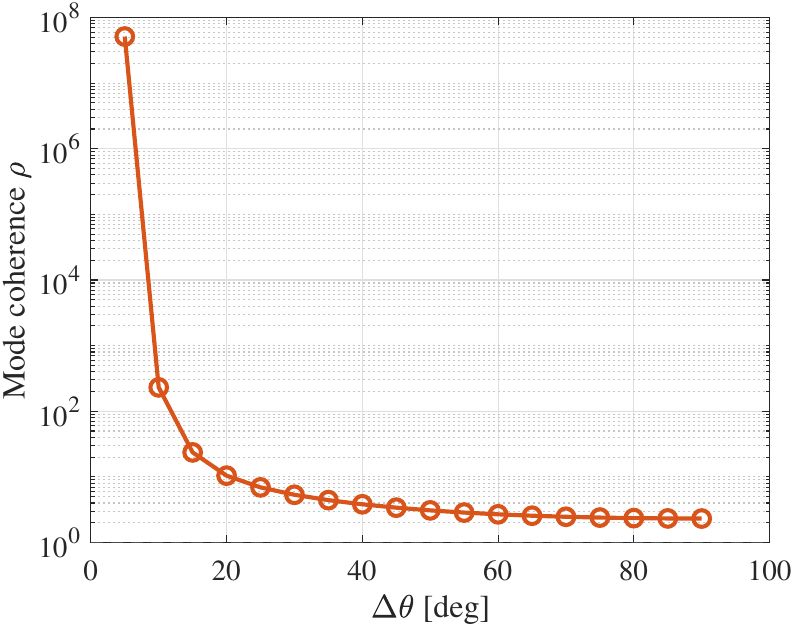}}
\subfigure[Cross-mode power and effective rank.]{\includegraphics[width=0.32\textwidth]{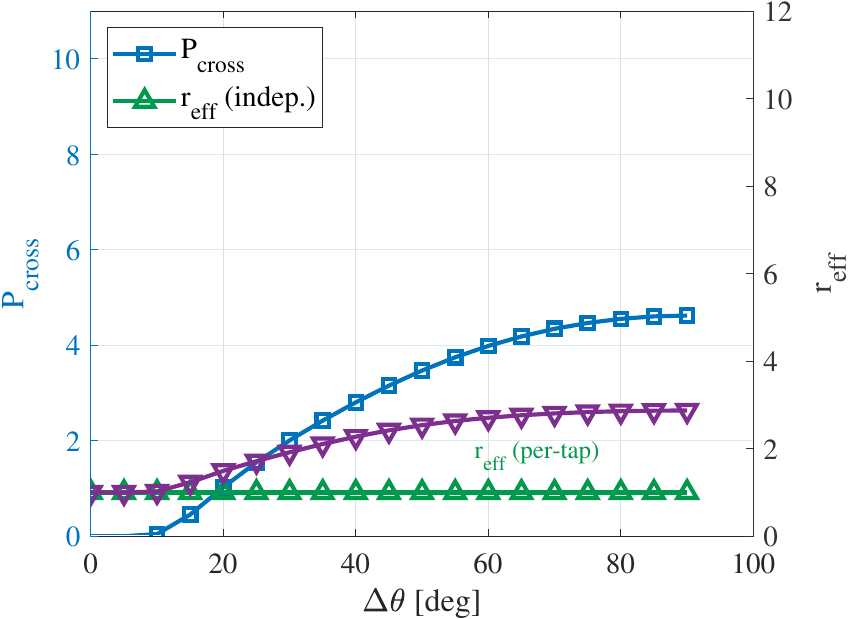}}
\subfigure[2D-Rake BER at $E_b/N_0=4$~dB.]{\includegraphics[width=0.32\textwidth]{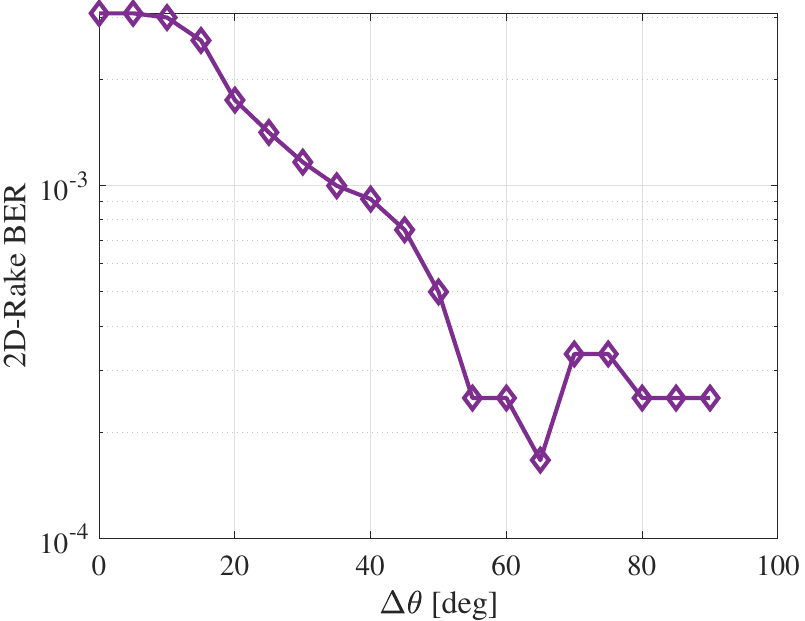}}
\caption{Mode coherence (a), cross-mode power and target-mode effective rank (b), and 2D-Rake BER (c) versus $\Delta\theta$.}
\label{fig:rank}
\end{figure*}

Figure~\ref{fig:rank} addresses the central question of Section~\ref{sec:divD} directly: \emph{when does multipath-induced modal leakage become recoverable diversity?} Panel (a) shows $\rho$ decreasing to $\approx2.3$ at $90^{\circ}$; panel (b) shows the cross-mode power $P_{\mathrm{cross}}$ growing by more than an order of magnitude over the same range. The decisive quantity is the effective rank $r_{\mathrm{eff}}$ \eqref{eq:reff} of the target-mode modal covariance $\mathbf{R}_{k,\ell}$: for the \emph{per-tap fading} channel (Section~\ref{sec:fading}) it stays at $1$ over the whole range, so the leaked energy is added but not independent and does not by itself create modal diversity; for the \emph{independent-branch} channel it grows from $1$ to $\approx2.9$; it would approach the $L=4$ limit only if all $L$ modal branches carried equal power. Panel (c) shows the 2D-Rake BER nevertheless improving everywhere: the receiver harvests \emph{energy} even when the \emph{rank} is low, which is the precise sense in which ``more IMI'' and ``more diversity'' differ: leakage provides energy unconditionally, and independent diversity only as far as the covariance rank allows.

\subsection{Imperfect CSI}
\label{sec:csi}
\begin{figure}[t]
\centering
\subfigure[Relative CSI error.]{\includegraphics[width=0.47\columnwidth]{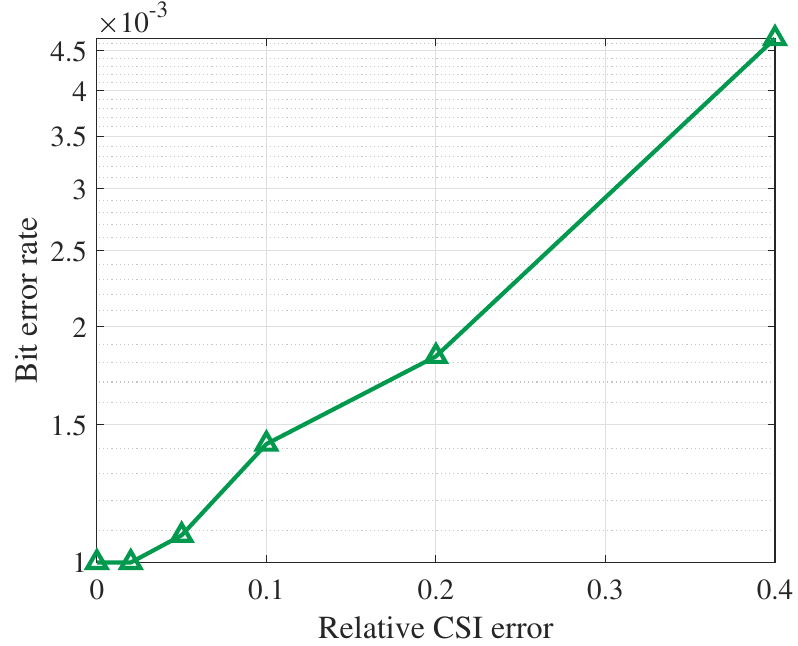}}
\subfigure[Pilot-based least-squares CSI.]{\includegraphics[width=0.47\columnwidth]{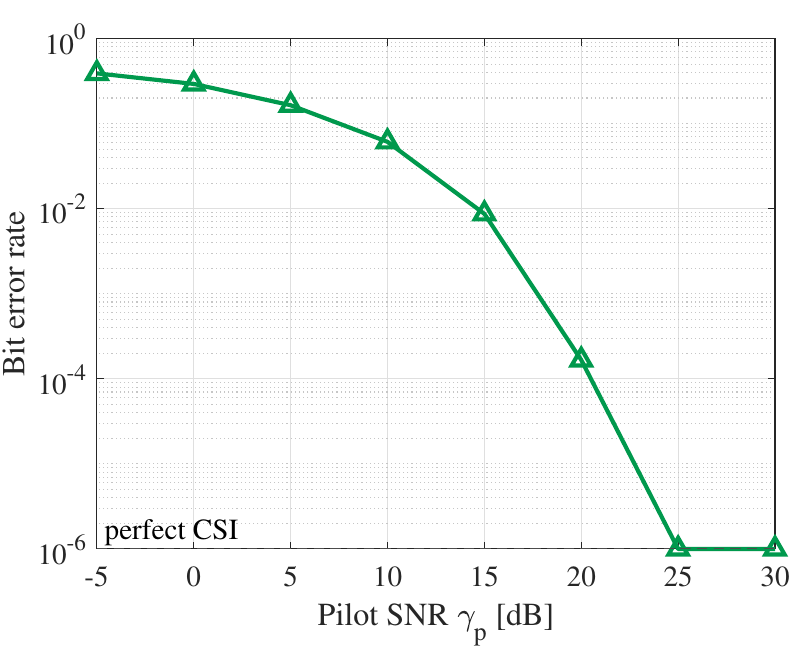}}
\caption{2D-Rake robustness to channel-estimation error (a) and to the pilot SNR (b).}
\label{fig:csi}
\end{figure}

Figure~\ref{fig:csi}(a) perturbs, at $E_b/N_0=4$~dB, the receiver's channel estimate by a relative error of up to $40\%$: the BER grows from $10^{-3}$ to $4.7\times10^{-3}$, less than a factor of five. Figure~\ref{fig:csi}(b) links the pilot budget to the BER through the least-squares model \eqref{eq:ls}: at a data $E_b/N_0=8$~dB the 2D-Rake needs a pilot SNR $\gamma_p\approx20$~dB to match perfect-CSI performance, degrades to $6.1\times10^{-2}$ at $\gamma_p=10$~dB, and is interference-dominated below $\gamma_p=0$~dB.

\subsection{Timing-offset robustness}
\label{sec:timing}
\begin{figure}[t]
\centering
\includegraphics[width=0.82\columnwidth]{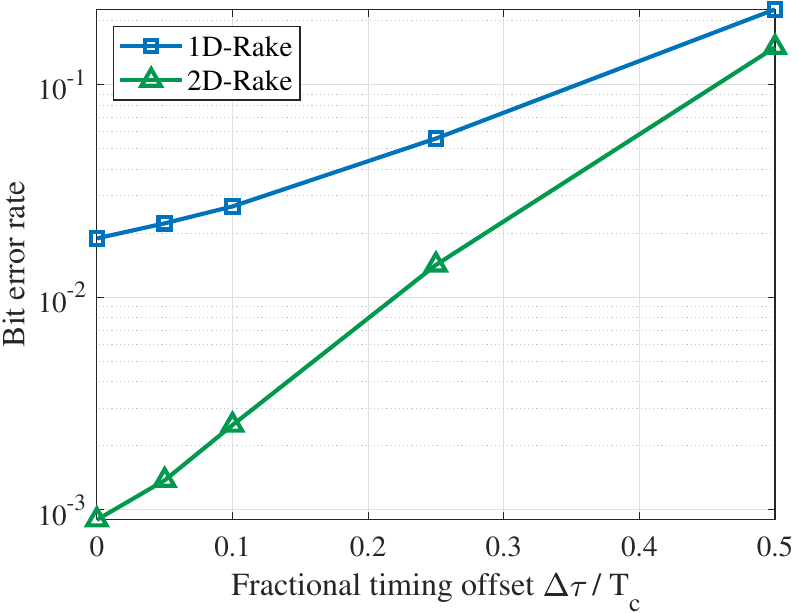}
\caption{BER versus the fractional chip-timing offset at $E_b/N_0=4$~dB.}
\label{fig:timing}
\end{figure}

The coloring design rests on the zero-lag code orthogonality \eqref{eq:orthcode}, which assumes chip-level timing synchronization. Figure~\ref{fig:timing} stresses this assumption with a common fractional offset $\Delta\tau/T_c$ between the transmit chip clock and the receive sampling grid (the receiver is unaware of the offset and uses the nominal integer-delay model; linear chip interpolation approximates the waveform). The 2D-Rake BER grows from $9.0\times10^{-4}$ at $\Delta\tau=0$ to $2.5\times10^{-3}$ at $0.1T_c$, $1.4\times10^{-2}$ at $0.25T_c$, and $1.5\times10^{-1}$ at $0.5T_c$ (the 1D-Rake: $1.9\times10^{-2}\to2.3\times10^{-1}$): the scheme is robust to sub-0.1-chip offsets and degrades gracefully beyond, but timing synchronization is a genuine structural requirement of TDOCC.

\begin{table*}[b]
\centering\footnotesize\setlength{\tabcolsep}{4pt}
\caption{Qualitative comparison with prior OAM multipath/combining schemes.}
\label{tab:prior}
\begin{tabular}{@{}l c c c c c c@{}}
\toprule
Scheme & Spread. & Delay-Rake & Cross-mode harv. & Joint IMI/ISI suppr. & Wireless & Ref.\\
\midrule
OAM-MDM (DFT demux) & -- & -- & -- & -- & yes & \cite{chen2020multimode}\\
OAM+OFDM phase compensation & -- & -- & -- & partial (sparse) & yes & \cite{liang2020joint}\\
MIMO equalization & -- & -- & partial & partial & yes & \cite{icoch2016mimo}\\
OAM-CDMA in free-space optics & yes & -- & -- & -- & optical & \cite{singh2022oamocdma}\\
Spread-spectrum OAM (security) & yes & -- & -- & -- & yes & \cite{zhang2020spread}\\
Fiber OAM-MDM MIMO-DSP & -- & -- & partial & partial & optical & \cite{bozinovic2013terabit}\\
\textbf{This work (2D-Rake)} & \textbf{yes} & \textbf{yes} & \textbf{yes} & \textbf{yes} & \textbf{yes} & --\\
\bottomrule
\end{tabular}
\end{table*}

\subsection{Fair comparison at equal energy per information bit}
\begin{figure}[t]
\centering
\includegraphics[width=0.82\columnwidth]{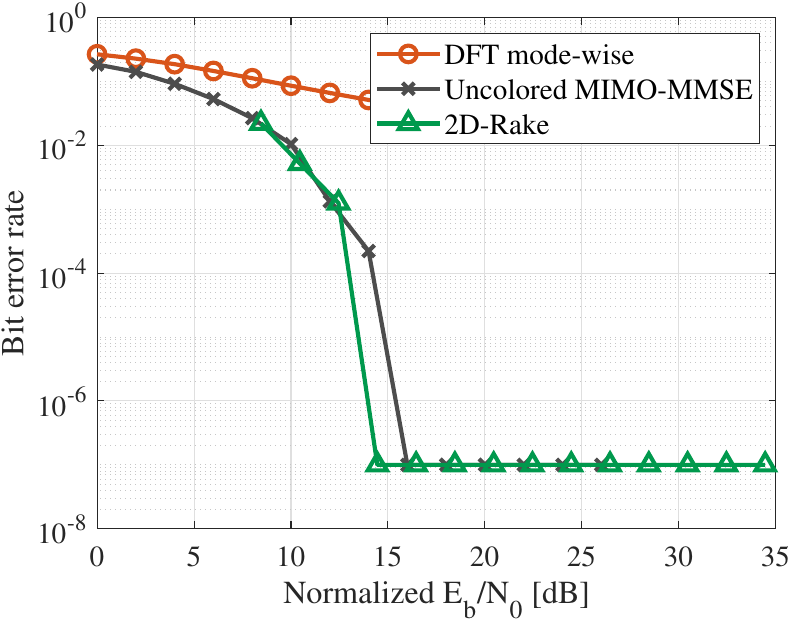}
\caption{BER versus normalized $E_b/N_0$.}
\label{fig:se}
\end{figure}

Figure~\ref{fig:se} re-plots the core comparison under the normalization of Section~\ref{sec:divE} (the colored curve shifted right by $10\log_{10}(1/\eta)\approx8.5$~dB for $N_c=4$, $N_g=3$). After the shift, the 2D-Rake crosses $10^{-3}$ at $\approx12.5$~dB normalized $E_b/N_0$, while the mode-wise detector is pinned above $1.5\times10^{-2}$ at every SNR. Against the uncolored modal-time MMSE equalizer, the strongest conventional receiver, the comparison at equal energy per bit is close: that equalizer crosses the same target at $\approx12.7$~dB normalized, yet the 2D-Rake stays ahead at every normalized $E_b/N_0$, by about $1.6$ times at the crossing and by up to an order of magnitude where the colored structure matters most. Against the delay-collapsed mode-domain MMSE detector the advantage is larger ($\approx12$~dB for $2\times10^{-2}$ versus $\approx8.5$~dB normalized, about $\mathbf{3.5}$~dB better).

\subsection{Spectral-efficiency trade-off}
\label{sec:setradeoff}
\begin{figure}[t]
\centering
\subfigure[Required normalized $E_b/N_0$.]{\includegraphics[width=0.47\columnwidth]{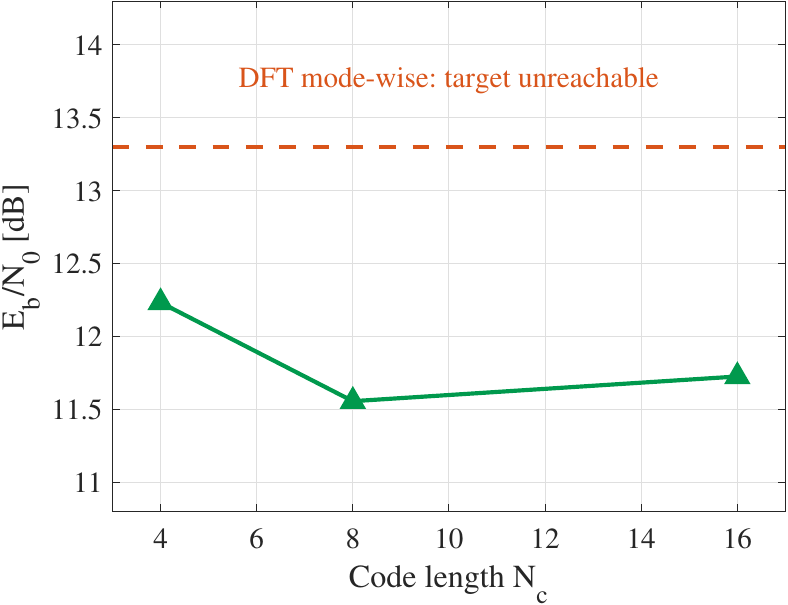}}
\subfigure[Spectral efficiency.]{\includegraphics[width=0.47\columnwidth]{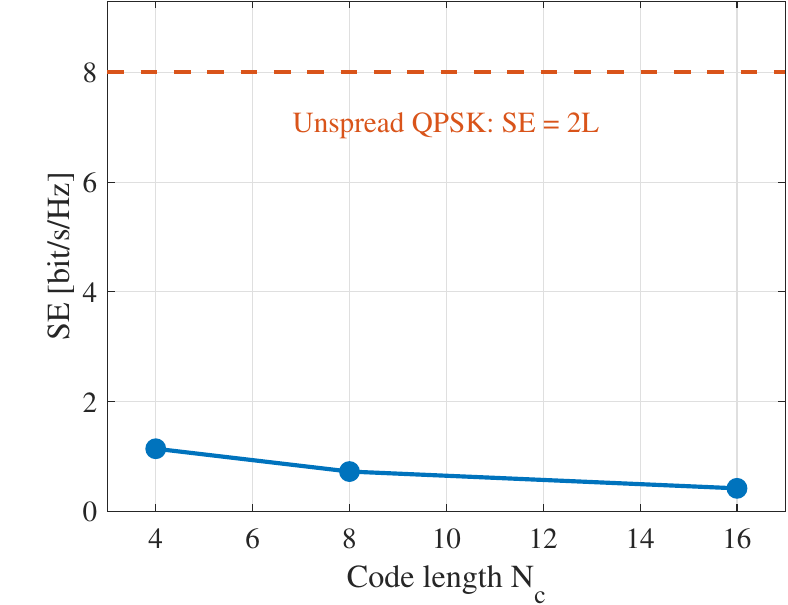}}
\caption{Spectral-efficiency trade-off at a fixed target BER.}
\label{fig:setradeoff}
\end{figure}

Figure~\ref{fig:setradeoff} quantifies the trade-off of Section~\ref{sec:divE} across the code-length axis: the required normalized $E_b/N_0$ falls from $\approx12.2$~dB ($N_c=4$) to $\approx11.6$~dB ($N_c=8$) as the longer code deepens the processing gain, then rises to $\approx11.7$~dB ($N_c=16$) as the overhead $(N_c+N_g)$ consumes the added gain, exposing a sweet spot. The spectral efficiency, meanwhile, drops monotonically from $1.14$ bit/s/Hz ($N_c=4$) to $0.42$ bit/s/Hz ($N_c=16$) against the $8$ bit/s/Hz of the unspread QPSK system, whereas the DFT mode-wise detector cannot reach the $10^{-3}$ target at any SNR (floor $\approx1.8\times10^{-2}$). The coloring thus turns an unreachable operating point into a reachable one, at a quantified rate cost, complementing the fixed-$N_c$ comparison of Fig.~\ref{fig:se}.

\subsection{Comparison with prior art and discussion}

Table~\ref{tab:prior} situates the proposal against the prior art: each ingredient has precedents, but the combination is distinct, since per-mode \emph{time-domain} orthogonal spreading separates the modes at zero delay and the joint mode$\times$delay Rake converts part of the multipath-induced leakage into a recoverable diversity resource under sufficiently separable conditions (Section~\ref{sec:divD}) instead of discarding it.

Because the MMSE combiner orthogonalizes the residual inter-code interference, Walsh--Hadamard and Zadoff--Chu coloring give essentially identical 2D-Rake BER ($2.15\times10^{-2}$ vs.\ $1.92\times10^{-2}$ at 0~dB); the family matters only for the MRC Rake.

Two practical remarks close this section. \emph{(i) Complementarity with beam design.} The receiver acts on whatever wavefront survives the channel, so it composes with transmit-side beam shaping \cite{wang2024constantradius, cheng2024fractal}: shaping raises the mode coherence $\rho$ and the 2D-Rake harvests the residual, so the two mechanisms multiply rather than compete. \emph{(ii) Multi-user extension.} A second tier of user-specific scrambling turns the scheme into OAM-CDMA multiple access, with the 2D-Rake extended to a three-dimensional (user$\times$mode$\times$delay) finger grid.

\section{Conclusions}
\label{sec:conclusion}

We have adapted the CDMA two-dimensional Rake principle to the OAM modal domain, to recover the residual modal orthogonality that conventional OAM demultiplexing discards in multipath: each OAM mode is colored with an orthogonal time-domain spreading code, and an $L\times K$ mode-delay finger grid jointly combines the despread outputs, harvesting the energy scattered across modes and delays. Built on the Bessel order-difference coupling law and a Gaussian mode-diffusion model validated against free-space field simulation, the analysis separated the asymptotic diversity order ($LK$ under independent-branch conditions, bounded by the modal covariance rank in general) from the finite-SNR effective diversity metric $D_{\mathrm{eff}}$, and the rank analysis showed that the adopted per-tap channel supports energy harvesting rather than independent modal diversity. At 28~GHz the 2D-Rake outperforms mode-wise, mode-domain ZF/MMSE, uncolored modal-time MMSE, and time-only Rake processing across the parameter space, stays robust to channel-estimation error, degrades gracefully under fractional timing offsets, and keeps its advantage after the spectral-efficiency cost of spreading is charged. The study uses the Gaussian mode-diffusion model (validated to $3.5\%$--$5.5\%$ NMSE but not an exact law) and a single compact configuration; broader sweeps of the array geometry, the Rician factor, and the code family remain, and validation against measured or ray-traced OAM channels, e.g., the geometry-based stochastic framework of \cite{lyu2025gbsm}, would strengthen the physical grounding. The spectral-efficiency trade-off \eqref{eq:se} invites joint optimization of the code length, the guard, and the modulation order; adaptive coloring, which switches the spreading on only when the channel is rich, is a natural extension, as are multi-user OAM-CDMA and a three-dimensional mode$\times$frequency$\times$time Rake.

\bibliographystyle{IEEEtran}
\bibliography{references}

\end{document}